\documentclass[fleqn,usenatbib]{mnras}

\usepackage{newtxtext,newtxmath}

\usepackage[T1]{fontenc}

\DeclareRobustCommand{\VAN}[3]{#2}
\let\VANthebibliography\thebibliography
\def\thebibliography{\DeclareRobustCommand{\VAN}[3]{##3}\VANthebibliography}

\usepackage{graphicx}	
\usepackage{amsmath}	
\usepackage{longtable}

\title[Eclipses in 47 Tuc O]{Study of consecutive eclipses of pulsar J0024$-$7204O}

\author[F. Abbate et al.]
{\parbox{\textwidth}{
F.~Abbate,$^{1,2}$\thanks{E-mail: federico.abbate@inaf.it}
A.~Possenti,$^{1}$
A.~Ridolfi,$^{1,2}$
S.~Buchner,$^{3}$
M.~Geyer,$^{3,4}$
M.~Kramer,$^{2,5}$
L.~Zhang,$^{6,7}$
A.~Corongiu,$^{1}$
F.~Camilo$^{3}$ and 
M.~Bailes$^{7,8}$
}
\\
\\
$^{1}$INAF -- Osservatorio Astronomico di Cagliari, Via della Scienza 5, I-09047 Selargius (CA), Italy\\
$^{2}$Max-Planck-Institut f\"{u}r Radioastronomie, Auf dem H\"{u}gel 69, D-53121 Bonn, Germany\\
$^{3}$South African Radio Astronomy Observatory, 2 Fir Street,Cape Town, 7925, South Africa\\
$^{4}$High Energy Physics, Cosmology and Astrophysics Theory (HEPCAT) Group, Department of Mathematics and Applied Mathematics, \\ University of Cape Town, Rondebosch 7701, South Africa\\
$^{5}$Jodrell Bank Centre for Astrophysics, Department of Physics and Astronomy, The University of Manchester, Manchester M13 9PL, UK\\
$^{6}$ National Astronomical Observatories, Chinese Academy of Sciences, A20 Datun Road, Chaoyang District, Beijing 100101,\\ People’s Republic of China\\
$^{7}$Centre for Astrophysics and Supercomputing, Swinburne University of Technology, Mail H39, PO Box 218, VIC 3122, Australia\\
$^{8}$ ARC Centre of Excellence for Gravitational Wave Discovery (OzGrav), Mail H29, Swinburne University of Technology, PO Box 218, Hawthorn, VIC 3122, Australia
}

\date{Accepted XXX. Received YYY; in original form ZZZ}

\pubyear{2024}

\begin{document}
\label{firstpage}
\pagerange{\pageref{firstpage}--\pageref{lastpage}}
\maketitle

\begin{abstract}
The eclipses seen in the radio emission of some pulsars can be invaluable to study the properties of the material from the companion stripped away by the pulsar. We present a study of six consecutive eclipses of PSR J0024-7204O in the globular cluster 47 Tucanae as seen by the MeerKAT radio telescope in the UHF (544-1088 MHz) band. 
A high scintillation state boosted the signal during one of the orbits and allowed a detailed study of the eclipse properties. We measure significant dispersion measure (DM) variations and detect strong scattering that seems to be the dominating mechanism of the eclipses at these frequencies. A complete drop in the linear polarization together with a small increase in the rotation measure suggests the presence of a magnetic field of $\sim 2$ mG. The study of multiple eclipses allowed us to measure difference in the lengths of the eclipses and DM differences of $\sim 0.01$ pc cm$^{-3}$ in consecutive orbits. One orbit in particular shows a delay in recovery of the linear polarization and a visible delay in the arrival of the pulses caused by a stronger scattering event. We suggest that these are caused by a higher variance of density fluctuations during the event.
\end{abstract}

\begin{keywords}
(Galaxy:) globular clusters: individual: 47 Tucanae -- (stars:) binaries: eclipsing -- scattering -- magnetic fields
\end{keywords}



\section{Introduction}
\label{intro}

Eclipsing pulsars are binary pulsars that are not visible or show significant distortions in the pulsed emission for sections of the orbital period. After the discovery of the first eclipsing pulsar PSR B1957+20 \citep{Fruchter1988} more and more pulsars have been found belonging to this class. 

The eclipses occur, with a few exceptions (e.g. PSR B1259$-$63 \citealt{Johnston1996} and PSR J0737$-$3039A \citealt{Lyne2004}) in a class of binary pulsars called `spiders'. These are systems with short orbital periods ($P_b<\sim1$d) and a non-degenerate companion star that is losing matter in the binary system \citep{Roberts2013} and are usually subdivided into ``black widows" and ``redbacks". The former have companion masses between $\sim 0.01$-0.05 M$_{\odot}$  while the latter has companion masses in the range $\sim 0.2$-0.4 M$_{\odot}$ \citep{Roberts2013}. The material lost by the companion is considered to be responsible for the distortion or the complete absorption of the pulsar emission. This is evidenced by the excess in dispersion measure (DM) often seen during the ingress and egress phases of the eclipses or when the pulsar signal is distorted.

The eclipses can occur at irregular orbital phases like in the case of PSR J1748$-$2446A \citep{Lyne1990}, but  are typically seen around the superior conjunction of the pulsar, when the latter is behind the companion and the apparent angular separation between the two is smallest. In many cases, the eclipse transition regions are asymmetrical with the egress region lasting longer than the ingress due to a comet-like tail of the material expelled by the companion \citep{Fruchter1988,Tavani1991}.
Also, the eclipses are often characterized by strong radio frequency dependence, lasting longer at lower frequencies and, in some cases, showing no signs of eclipses at high frequencies \citep{Stappers2001,Bhattacharyya2013,You2018,Polzin2019}. 

The mechanisms that cause the eclipses are not fully understood, with various ideas being proposed (e.g. \citealt{Thompson1994}) soon after the discovery of the first eclipsing systems. Among these are plasma frequency cut-off, free-free absorption, refraction of the anisotropic radio beam, and smearing of the pulsations due to the strong dispersion associated with the plasma released by the companion star. More recently, evidence has been collected that suggests, at least in some cases, the phenomenology in the egress region could also be dominated by the smearing of the pulsation due to scattering \citep{Polzin2020}. Other studies also point to cyclotron/synchrotron absorption as the main cause of the eclipse \citep{Kudale2020,Kansabanik2021}. This requires that magnetic fields of $\sim 10$ G are present at the very center of the eclipses. The magnetic fields have also been proposed as the main cause of the drop in linear polarization seen at the edge of the eclipses \citep{You2018,Polzin2019} and the change in circular polarization described by Faraday conversion \citep{Li2023}.

The pulse profile of some eclipsing pulsars shows strong scattering tails caused by the eclipsing material interposed along the line of sight \citep{Johnston1996,Bai2022}. The tail can be parameterized by a scattering time $\tau$ with a power-law dependence on the observing frequency. The expected value for the spectral index $\alpha$, called scattering index, in case of an isotropic Gaussian electron density spectrum  is $-4$ \citep{Cronyn1970,Lang1971} while for a Kolmogorov spectrum of turbulence in the material the expected value is $-4.4$ \citep{Lee1976,Rickett1977}. In the general non-eclipsing pulsar population $\alpha$ scattering is also observed and $\alpha$ varies greatly from $-5.6$ to $-1.5$ \citep{Lewandowski2013,Lewandowski2015,Geyer2017}. Similar ranges of values, from $-4.6$ to $-1.6$, are measured also in eclipsing pulsars \citep{Bai2022}. 

Since the earliest papers on eclipsing pulsars \citep{Fruchter1988,Lyne1990}, significant variations in flux, time delays and changes in DM have been detected from eclipse to eclipse. These variations have been shown to occur on the timescales of a few days to a few weeks \citep{Polzin2020}. The few cases where two consecutive eclipses have been studied show that significant variations can occur even on the timescale of a single orbit \citep{Li2023,Kumari2023}. 

PSR J0024$-$7204O \citep[hereafter 47 Tuc O]{Camilo2000} is a 2.64 ms pulsar in the globular cluster 47 Tucanae (hereafter 47 Tuc). It is in a binary system with an orbital period of 3.26 h and a minimum companion mass of 0.025 M$_{\odot}$ \citep{Freire2017}. The occurrence of eclipses of the radio pulsations, which were first described by \cite{Freire2003}, led to the classification of this pulsar as a black widow system. The availability of an uninterrupted $\sim$17 h observation of 47 Tuc led to the opportunity of studying the pulsar flux density, DM, scattering and polarization variability over six consecutive eclipses, which is the focus of our work reported in this paper.

The paper is organized as follows. In \S \ref{sec:obs} we present the observations and describe the data analysis. In \S \ref{sec:eclipse_properties} we report on the properties of the brightest eclipse while the variability of the eclipses is addressed in \S \ref{sec:eclipse_variability}. In \S \ref{sec:discussion} we analyze the consequences regarding the magnetic field, mass loss and scattering index.

\section{Observations and analysis}\label{sec:obs}

The globular cluster 47 Tuc was the target of an intensive observing campaign at the MeerKAT radio telescope in the context of the MeerTime\footnote{\url{http://www.meertime.org}} \citep{Bailes2016, Bailes2020} and TRAPUM (TRansients And PUlsars with MeerKAT \footnote{\url{http://www.trapum.org}}) \citep{Stappers2016} Large Survey Projects (LSPs). The campaign consisted of 23h of observations from 2022 Jan 26 to Jan 29 and is described in detail in \cite{Abbate2023}. The duration of the longest observation was 17 h, enough to cover six consecutive orbits (and therefore eclipses) of 47 Tuc O. The egress of the last eclipse happened after the end of the observation.

For the purpose of this paper we only looked at the 17h observation made on 2022 Jan 27 at the UHF band over the frequency range 544-1088 MHz. The observation made use of the Pulsar Timing User Supplied Equipment (PTUSE) machines \citep{Bailes2020} which have the ability to synthesize four different tied-array beams to observe different regions of the cluster simultaneously. One of the beams was aimed at the optical centre of the GC, very close to where 47 Tuc O is located. 
The position and size of the beams used in this observation are shown in Figure 1 of \cite{Abbate2023}. 

The beam was formed using 50 antennas and data was acquired in search mode with full polarization and a sampling time of 7.5 $\mu$s. The observation was recorded with 4096 frequency channels, coherently de-dispersed within each channel and incoherently between the channels at a value of 24.404  pc cm$^{-3}$ (chosen as the average of the DMs of the pulsars within the beam). The number of frequency channels was later reduced by a factor of 16 and only 256 were stored for future analysis.

Before each observation, the MeerKAT array goes through a phase-up and calibration procedure in order to obtain both the correct delays from each antenna and polarization calibration \citep{Serylak2021}. That is achieved by observing a well-known calibrator (either PKS J0408$-$6546, PKS J0825$-$5010 or PKS J1939$-$6342, depending on which one is visible during the observation) and noise diodes. The process lasts around 15 minutes and, in order to maintain a good solution, must be repeated after at most 5 hours. During the 17 h observation, the calibration was repeated 5 times leaving some gaps in the data. The phase-up times were chosen to be made at orbital phases far from the eclipse of 47 Tuc O.  

The search mode data of the beam containing 47 Tuc O were folded using the \texttt{DSPSR}\footnote{{\url{http://dspsr.sourceforge.net}}} pulsar package \citep{vanStraten2011}. After the folding, we used the routine \texttt{pac} from \texttt{PSRCHIVE}\footnote{\url{http://psrchive.sourceforge.net}} \citep{Hotan2004,vanStraten2012} to correct for the variations of the parallactic angle. About 8 percent of the channels in the UHF band at MeerKAT were affected by radio frequency interference (RFI). We removed the affected channels using the \texttt{pazi} routine of \texttt{PSRCHIVE}.

\begin{figure*}
\centering
	\includegraphics[width=\textwidth]{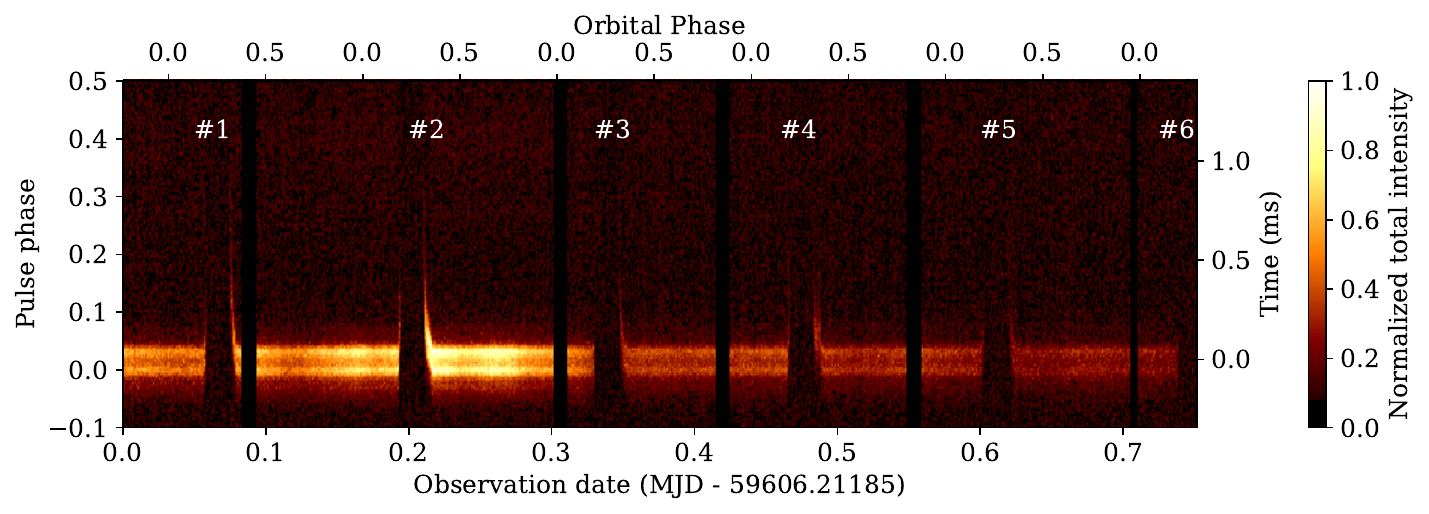}
    \caption{Plot of the pulse profile of 47 Tuc O across the entire 17 h observation. The pulse profile is plotted as a function of pulse phase (left) and time (right). The black vertical stripes show the time and duration of the phase-up calibrations done throughout the observation. A total of 6 eclipses (6 ingresses and 5 egresses) can be seen. A small delay can be seen immediately before the eclipses, while a longer tail with significant pulse broadening can be seen as the pulsar exits the eclipses. The color scale shows the total intensity normalized to arbitrary units. The flux variations during the observation are caused by the ISM scintillation that affects all pulsars in 47 Tuc. The sub-integration time (i.e. the temporal resolution along the horizontal axis) is 64s.}
  	\label{fig:eclipse_flux}
\end{figure*}

Like many other black widows, 47 Tuc O shows strong orbital variability \citep{Freire2017} which can be hard to predict and can cause significant drifts in the folded profile. To solve this problem, we derived a local timing solution masking the orbital phases affected by the eclipse as described in \cite{Abbate2023}. The resulting pulse profile shows no trends in pulse phase as a function of time with the exceptions of the extra delays caused by the eclipse as can be seen in
Figure \ref{fig:eclipse_flux}. The strong flux variations seen in the Figure are caused by scintillation due to the interstellar medium (ISM) along the line of sight between the Earth and the pulsar.



To allow for precise measurements of the eclipse properties we averaged over 32s of data for the eclipse \#2 which happened during a peak of scintillation and over 64s of data for the other eclipses.
We determined the value of DM and scattering time for each sub-integration using the code \texttt{PulsePortraiture}\footnote{\url{https://github.com/pennucci/PulsePortraiture}}\citep{Pennucci2014,Pennucci2019}. In order to increase the signal to noise ratio (S/N) in each frequency channel, we reduced the number of channels to 32 for this last step. The scattering time, if not properly corrected, can bias the determination of the DM. We fixed this in \texttt{PulsePortraiture} with an option that simultaneously fits for DM and scattering time. In particular, the scattering time is fitted assuming an isotropic scattering model.
After correcting each sub-integration we the value of measured DM, we estimated the total flux, linear polarization percentage, circular polarization, the absolute value of the circular polarization percentage, as described in \cite{Tiburzi2013}. For each sub-integration we also determined the Faraday rotation measure (RM) by using the \texttt{rmfit} routine of \texttt{PSRCHIVE}. The ionosphere causes daily variations of the order of 0.5-1 rad m$^{-2}$ \citep{Porayko2019}. We accounted for this effect using the software \texttt{RMextract}\footnote{\url{https://github.com/lofar-astron/RMextract}} \citep{Mevius2018}. This piece of software estimates the ionospheric RM at a certain position in the sky and time by using a geomagnetic field model, the World Magnetic Model\footnote{\url{https://www.ngdc.noaa.gov/geomag/WMM/DoDWMM.shtml}}, and a global ionospheric map built from Global Navigation Satellite Systems (GNSS) data, the Center for Orbit Determination in Europe global ionospheric map  (CODG)\footnote{\url{https://www.aiub.unibe.ch/research/code___analysis_center/index_eng.html}}.

\section{Properties of the best measured eclipse} \label{sec:eclipse_properties}

During the eclipse \#2 the signal was amplified by a peak of scintillation caused by the ISM as can be seen in Fig. \ref{fig:eclipse_flux}. This allows us to study the properties of this eclipse in greater detail.


\begin{figure}
\centering
	\includegraphics[width=0.49\textwidth]{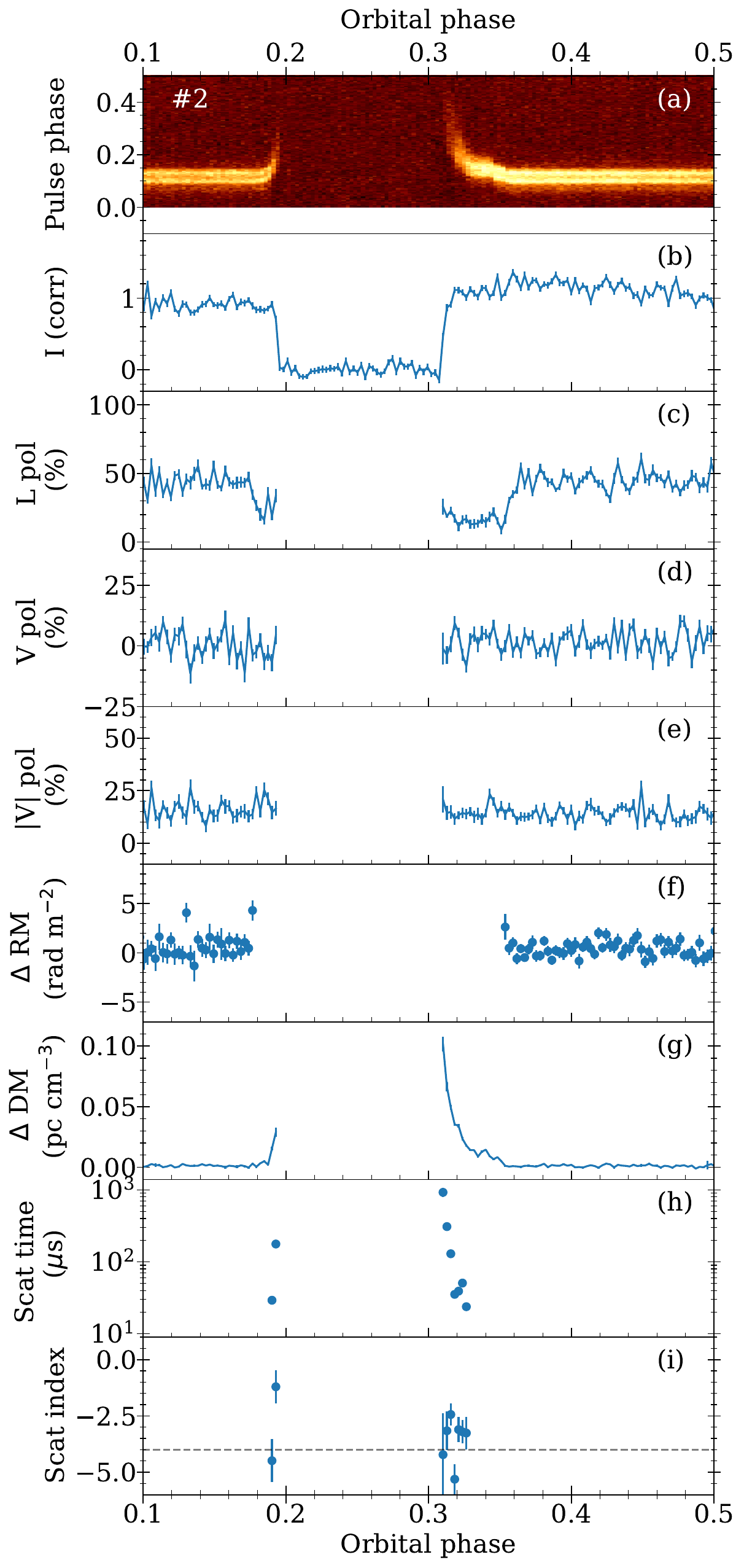}

    \caption{Properties of eclipse \#2 observed during the 17 h observation of 47 Tuc O. In the upper panel (a), we report the total intensity as a function of pulse phase and orbital phase. Below that panel we show the integrated total intensity flux in arbitrary units (b), normalized so that the average flux is 1 in the range of orbital phase between 0.05 and 0.15. The value of the flux after the end of the eclipse is higher than 1 due to the ISM scintillation. In the following panels we show the linear, L, (c) circular, V, (d) and absolute value, $|{\rm V}|$, of the circular polarization percentages (e), the RM (f) and the DM differences (g) from the value reported in {\protect \cite{Abbate2023}}, the scattering time at 800 MHz (h) and the scattering index (i). The RM excess are plotted only for the orbital phases where a clear detection was possible. The scattering time and scattering index are shown only for the sub-integrations where the fits returned a scattering time longer that 22.5 $\mu$s, which corresponds to 3 times the sampling time. In the panel of the scattering index, the dotted line shows the expected value for an isotropic Gaussian electron column density spectrum of $-$4. Each quantity has been measured as the integrated value over 32s sub-integrations. All the errors are at 1$\sigma$.}
  	\label{fig:26U2_full}
\end{figure}

In Figure \ref{fig:26U2_full} we show a zoom in of the plot in Figure \ref{fig:eclipse_flux} around the eclipse \# 2 together with the total flux, linear and circular polarization percentages and the absolute value of the circular polarization, RM and DM differences from the values of $+25.76$ rad m$^{-2}$ and $24.3580$ pc cm$^{-3}$ reported for 47 Tuc O in \cite{Abbate2023}, the scattering time at 800 MHz and the scattering index. 
Values for the linear, circular and absolute value of the circular polarization, and the DM excess have not been considered in the region where the total flux is less than 5 times the rms value of the noise. The RM excesses has been considered only for the orbital phases where the \texttt{rmfit} code is able to detect pulsations in the linear polarization profile, thanks to high S/N or high values of polarization. The scattering time and scattering index have been considered only for those sub-integrations where the fits returned a scattering time longer than 22.5 $\mu$s, which corresponds to 3 times the sampling time.
  

\subsection{Eclipse duration} \label{sec:eclipse_duration}

The evolution of the flux density across the eclipse can be approximated by a Fermi-Dirac function \citep{Broderick2016,Polzin2018}. We can use this function to derive the length of the eclipse and the length of the transition region. The Fermi-Dirac function takes the form 
\begin{equation}\label{eq:fermi-dirac}
    f= A(e^{\frac{2(\Phi-p_1)}{p_2}} +1)^{-1},
\end{equation}

where $A$ is the flux density out of the eclipse, $\Phi$ is the orbital phase, $p_1$ is the orbital phase at half maximum and $p_2$ is related to the width of the transition in units of orbital phase. The parameter $p_2$ is defined as the region in which the flux changes from a factor of $1/(e^{-1}+1) \sim 0.73$ to $1/(e^{1}+1) \sim 0.27$. For the ingress it is positive while for the egress it is negative. We perform separate fits during the ingress and egress using the Bayesian fitting algorithm \texttt{emcee} \citep{Foreman-mackey2013}. In order to account for the ISM scintillation we allow the parameter $A$ to change between the ingress and the egress. The length of the eclipse, $\Delta\Phi_{\rm ecl}$, is defined as the difference between the best-fitting values of $p_1$ for the egress and ingress and expressed in orbital phases, reported in Table \ref{tab:eclipse_lengths}. We find $\Delta\Phi_{\rm ecl} = 0.118 \pm 0.001$. All the errors throughout the paper, unless specified, are at the 1$\sigma$ level. As already shown in \cite{Freire2003}, the size of the eclipse is significantly larger than the Roche lobe of the companion. Therefore the material responsible for the eclipse is not gravitationally bound to the companion.

The best-fitting values of the width of the transition area, $p_2$
of the ingress and egress are, respectively, $0.0012 \pm 0.0007$ and $-0.0028 \pm 0.0002$ (shorter than one 32s sub-integration for the ingress and slightly longer for the egress) where the sign indicates if the flux is increasing or decreasing. This suggests that the absorption of the flux occurs slightly faster during the ingress probably due to a sharper rise of the density of the intervening medium. This can be confirmed by looking at Fig. \ref{fig:freq_dependence}. This figure shows a zoom in of the ingress and the egress with the integrated flux over the full band (black dashed line) and divided into four sub-bands. The frequency evolution of the eclipses is discussed in \S \ref{sec:Freq_dependence}.


\subsection{DM excess and scattering}

Despite the almost symmetrical behaviour of the total intensity integrated in frequency and pulse phase (black dashed line in Fig. \ref{fig:freq_dependence}), 
the pulse-phase resolved intensity (top plots of Fig. \ref{fig:freq_dependence}) looks strongly asymmetric. This is caused by an excess of DM as can be seen in panel (g) of Fig. \ref{fig:26U2_full}. During the ingress, significant DM differences from the value measured outside of the eclipse are visible in the last five sub-integrations ($\sim$ 0.01 orbital phases) before the pulsar becomes undetected. The highest measured value is $\Delta {\rm DM_{max,ing}}=0.029\pm0.004$ pc cm$^{-3},$ corresponding to an extra column density of ionized gas of $(9 \pm 1) \times 10^{16}$ cm$^{-2}$. Instead, during the egress, the DM goes back to the average value about 16 sub-integrations ($\sim$ 0.043 orbital phases) after the pulsar becomes visible again. The highest measured value is $\Delta {\rm DM_{max,eg}}=0.101\pm0.006$ pc cm$^{-3},$ corresponding to an extra column density of ionized gas of $(3.0 \pm 0.2) \times 10^{17}$ cm$^{-2}$.
This asymmetric behaviour of the DM excess can be explained by a comet-like tail in the material surrounding the companion.

The scattering time follows a similar trend to the DM. It starts growing at the same time as the DM and reaches a maximum value of $\sim 177$ $\mu$s in the ingress and a value of $\sim 930$ $\mu$s during the egress stage. In a single 32s-integration in the egress, the scattering time grows by a factor of three from $\sim 309$ $\mu$s to $\sim 930$ $\mu$s. 
If an increase of the same factor occurs in the preceding integration as well, the scattering time will be $\sim 2.7$ ms at 800 MHz, longer than the rotational period of the pulsar. This would wash out completely the pulsations and render the pulse indistinguishable from the noise. During the ingress, the increase in scattering time over the last 32s-integration where the pulsar is visible increases roughly by a factor of six. Therefore, at the frequencies covered by this observation, the size of the eclipse seems to be dominated by scattering, similar to the case of PSR J1816+4510 \citep{Polzin2020}. 

In the assumption that the scattering is the single cause for the observed eclipse, we can predict how the flux should change. In isotropic scattering no flux is lost but the pulsed flux can decrease for strong scattering. If the scattering tail is wider than the window where the pulsed flux is measured, only part of the flux will be counted as the pulse while the rest will contribute to raising the baseline. In our case only the first two sub-integrations after the egress have strong enough scattering to show this effect. To estimate how much flux is lost through scattering alone, we simulate a scattered profile with the same properties as measured in these two sub-integrations. The scattered profile is the convolution between the pulsar profile and the temporal broadening function due to isotropic scattering. In formulas it can be expressed as: $P_{\rm scat}(t) = P_{\rm int}(t) * f_{\rm scat}(t)$, where $P_{\rm int}(t)$ is the pulse profile unaffected by scattering and $f_{\rm scat}(t)$ is the temporal broadening function defined as:

\begin{equation}
    f_{\rm scat}(t)= \tau^{-1} e ^{-t/\tau} U(t),
\end{equation}
where $U(t)$ is the unit step function to ensure that we only consider the time $t>0$. For the reference pulse profile we take an average of the profile in the egress of eclipse \#2 (excluding the sub-integrations affected by scattering) divided in four frequency channels in order to take into account the effects of scintillation. We numerically convolve this profile with the temporal broadening function for each frequency using the measured values of scattering times, respectively $926 \pm 136$ $\mu$s and $309 \pm 26$ $\mu$s. The reported scattering times are measured at 800 MHz so we calculate the expected value at each frequency channel using the measured scattering indices of $-4.2 \pm 1.8$ and $-3.2 \pm 0.9$.
Once we obtain the scattered profiles we calculate the pulsed flux using the same method as for the observed sub-integrations. We repeat the calculation for 5000 times in order to minimize the effects of the noise and of the uncertainties on the scattering properties.  For the first sub-integration we obtain a flux of $0.56\pm 0.07$ compared to the measured value of $0.47 \pm 0.05$. For the second sub-integration we obtain a flux of $0.92 \pm 0.07$ compared to the measured value of $0.87 \pm 0.05$. In both cases the predictions match the observations at the 1$\sigma$ level showing that scattering can caused the observed decrease in flux.

The different values of the scattering time and DM excess during the ingress and egress stage can be explained by a much sharper increase of the density of the eclipsing material in the ingress. This is in line with the idea of a comet-like tail where the density of the material in the front has a sharper gradient than in the tail.

The scattering index during the ingress changes from $\sim -4.5$ to $\sim -1.2$ while during the egress stage is between $-5$ and $-2.5$. These variations might be caused by the varying geometry of the scattering screen in the eclipsing medium. More details are given in \S \ref{sec:scattering_index}. Changes of the same magnitude have also been noticed in PSR B1957+20 \citep{Bai2022}.



\subsection{Linear polarization}\label{sec:polarization}
Another important effect that we note is the effect of the eclipse on the linear polarization percentages (Figure \ref{fig:26U2_full}, panel c). The linear polarization shows a sharp decrease close to edges of the eclipse. The linear polarization drop occurs as soon as the DM starts to increase.  During the ingress the DM excess is $\sim 0.0033\pm 0.0007$ while during the egress the DM excess reaches $\sim 0.0051 \pm 0.0009$ pc cm$^{-3}$ as the linear polarization starts to drop. Similar effects have been seen in other eclipsing pulsars, like PSR J1748$-$2446A \citep{You2018} and PSR J2051$-$0827 \citep{Polzin2019}.

Simultaneously with the drop in linear polarization, we stop being able to measure the RM in a single sub-integration, due to the fact that the \texttt{rmfit} code becomes unable to detect any pulsation in the linear polarization profile. In turn, this means that we are not able to measure the component of the magnetic field parallel to the line of sight in the most central regions of the eclipse. However, just before disappearing, both in the ingress and the egress phases, the RM shows an increase from the average value. The RM excess of the last point in the ingress region is $4.3 \pm 1.0$ rad m$^{-2},$ while in the egress region is $2.6 \pm 1.3$ rad m$^{-2}$. This increase is not significant, being similar to the value measured at other orbital phases and being consistent with the average value at 4$\sigma$ for the ingress and at 2 $\sigma$ for the egress. However, the occurrence at both rims of the eclipse and right next to the drop in linear polarization suggests that the presence of a magnetic field in the eclipsing material might be the cause of the rise in RM. 
{In those sub-integrations the DM excess is measured to be $0.0030 \pm 0.0008$ pc cm$^{-3}$ and $0.0012 \pm 0.0007$ pc cm$^{-3},$} for the ingress and egress cases, respectively. In the hypothesis that the RM and DM excesses come from the same material and that the variations of electron density are uncorrelated to the variations of magnetic field, the average component of the magnetic field parallel to the line of sight can be measured using the equation:

\begin{equation} \label{eq:magnetic_field}
    \langle B_{\parallel}\rangle [\mu {\rm G}]\sim 1.23 \times {\rm \Delta RM [rad \,m^{-2}]}/{\rm \Delta DM [pc\, cm^{-3}]}.
\end{equation}

We find $1.8 \pm 0.6$ mG for the ingress region and $2.7 \pm 2.0$ mG for the egress region. In both cases, the average magnetic field is pointing towards the observer. These values are similar to the ones seen in the eclipsing pulsar PSR J2256$-$1024 \citep{Crowter2020}.

The drop in linear polarization is usually considered to be caused by rapid RM fluctuations within the region crossed by the line of sight during the time-span of the sub-integration. If we assume that the RM fluctuations a normal distribution with standard deviation $\sigma_{\rm RM}$, then the linear polarization percentage can be described by the equation \citep[eq. A5]{You2018}:

\begin{equation}
    L=L_0 e^{-(2\lambda^4 \sigma_{\rm RM}^2)},
\end{equation}
where $L_0$ is the linear polarization percentage outside the eclipse and $\lambda$ is the wavelength. For a wavelength of 0.37 m (corresponding to the central frequency of 816 MHz), we find that the linear polarization drops by a factor of $e$ if the $\sigma_{\rm RM}$ is $\sim 5$ rad m$^{-2}:$ if such a variability occurs within the same sub-integration, very little linear polarization would be seen. The variability can be due to fluctuations in DM or in the intensity or direction of the magnetic field. 
For a magnetic field with a magnitude of $\sim 2$ mG and a rapidly varying direction, the aforementioned $\sigma_{\rm RM}$ requires an extra DM of only 0.0025 pc cm$^{-3}$. This is compatible with the value of excess DM that is measured in the sub-integrations where the linear polarization drops. 
On the other hand, if the direction and intensity of the magnetic field are constant, the fluctuations of RM should be dominated by electron density variations within the same sub-integration. We can test the expected level of the magnetic field in this case. 
In correspondence with the linear polarization drop, the DM excess rises in a single 32s-subintegration from $0.0005 \pm 0.0010$ pc cm$^{-3}$ to $0.0033 \pm 0.0007$ pc cm$^{-3}$ during the ingress and from $0.0012 \pm 0.0007$ pc cm$^{-3}$ to $0.0051 \pm 0.0009$ pc cm$^{-3}$ during the egress. We can use the equation $\langle B_{\parallel}\rangle \sim 1.23 \,\sigma_{ \rm RM}/\sigma_{\rm DM}$ to estimate the value of the magnetic field necessary in order to have $\sigma_{ \rm RM} \sim 5$ rad m$^{-2}$ in the same subintegration. We obtain $\langle B_{\parallel}\rangle =2.2 \pm 1.3$ mG for the ingress and $\langle B_{\parallel}\rangle =1.6 \pm 0.7$ mG for the egress. These values are compatible with what estimated above.

\subsection{Circular polarization}

A small peak in the absolute value of the circular polarization is visible around the orbital phase 0.34 for the eclipses \#2.
A similar peak is seen also in eclipse \#1 at the same orbital phase and in eclipse \#4 at orbital phase 0.36 (see Fig. \ref{fig:26U13_full} and \ref{fig:26U45_full}). 
Interestingly in all three cases the peak seem to precede the drop of in the percentage of linear polarization by $\sim 0.01$ orbital phases. One possible way to explain this would be to invoke Faraday conversion as a mechanism to convert linear polarization into circular polarization \citep{Cohen1960, Zheleznyakov1964,Vedantham2019, Gruzinov2019}. This mechanism has been suggested as an explanation for the circular polarization of FRB 20201124A \citep{Xu2022} and detected in the eclipses of PSR B1744-24A \citep{Li2023}. However, this mechanism requires very large magnetic fields. Eq. 7 of \cite{Li2023} reports that the required magnetic field strength for Faraday conversion to occur is $B \gtrsim 1400 {\rm G} (f/{\rm 2 GHz}) \cos{\alpha_{\rm pitch}}$, where $\alpha_{\rm pitch}$ is the pitch angle between the wave vector and the magnetic field. This is almost six orders of magnitudes larger than the estimated values of the parallel component of the magnetic field at the edges of the eclipse of $\sim 2$mG. 

An alternative possibility that has been put forward by \cite{Beniamini2022} is multipath propagation through a magnetized scattering screen. This mechanism can induce circular polarization if the rotation measure induced by the screen, $\rm RM_s$ is (eq. (58 in \citealt{Beniamini2022})):

\begin{equation}
    {\rm RM_s} \gtrsim 10 \left( \frac{\nu}{\rm 1 GHz} \right)^2 {\rm rad \, m^{-2}}.
\end{equation}

Although we do not measure RM at these orbital phases, we can predict what the expected contribution of RM from the eclipsing medium would be by taking the value of the parallel component of the magnetic field of about 2 mG and the measured DM excess at the orbital phase of the peak, $0.0092\pm 0.0008$. Using eq. \ref{eq:magnetic_field} we obtain a value of RM of $\sim 15$ rad m$^{-2}$. This value is larger than what needed to induce circular polarization by multipath propagation.

\subsection{Frequency dependence}\label{sec:Freq_dependence} 

\begin{figure*}
\centering
	\includegraphics[width=1\textwidth]{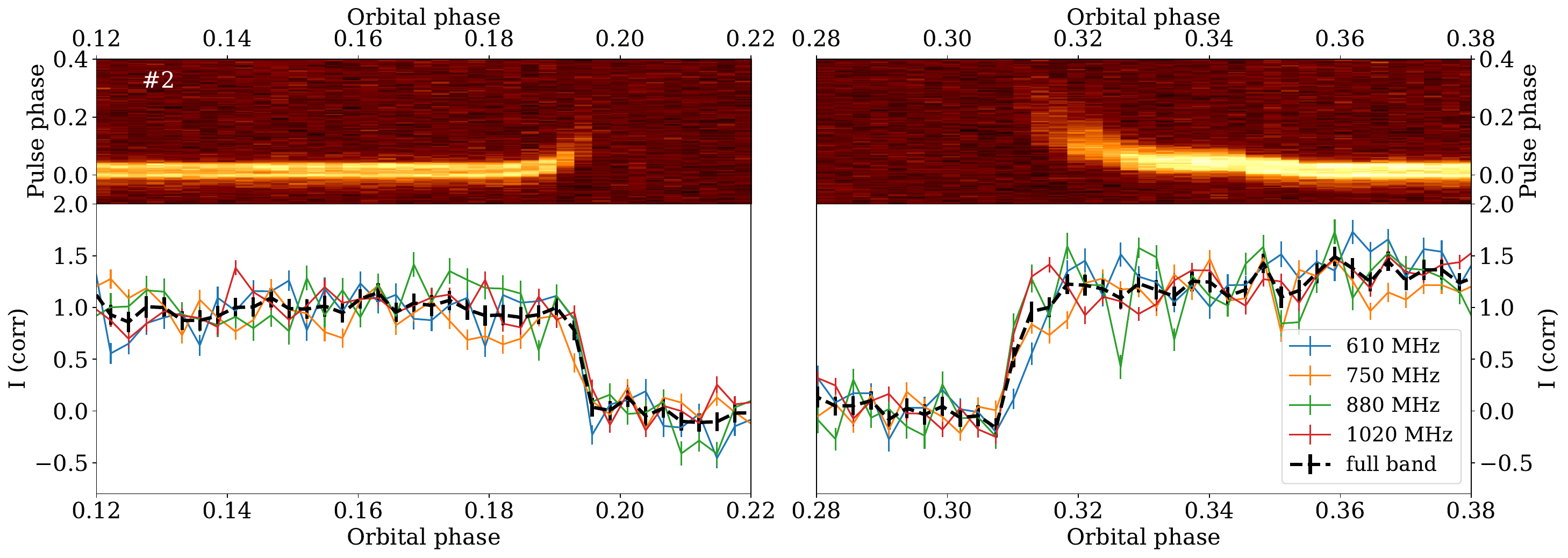}
    \caption{Plot showing the total intensity as a function of pulse phase and orbital phase (top panels) and the integrated total intensity (bottom panels) during the ingress (left panels) and the egress (right panels) of eclipse \#2. 
    The bottom panels show the integrated total intensity in the full band (black dashed line) and in four different sub-bands with central frequency reported in the legend. In the egress phase the eclipse lasts visibly longer at low frequencies compared to the higher frequencies. All the errors are at 1$\sigma$.}
  	\label{fig:freq_dependence}
\end{figure*}

Thanks to the large bandwidth of the observation (544-1088 MHz), we can divide the frequency band into four channels and calculate the size of the eclipse at the different channels by fitting with the same Fermi-Dirac function as done in \S \ref{sec:eclipse_duration}. The integrated total intensity in the four frequency bands at the ingress and egress can be seen in Fig. \ref{fig:freq_dependence}. The size of eclipse in units of orbital phases changes from $0.120\pm 0.002$ at 0.6 GHz to $0.115\pm 0.002$ at 1.0 GHz. We fit these values with a power-law $\Delta\Phi_{\rm ecl}(\nu)=\Delta\Phi_{\rm ecl, \nu_0} (\nu/\nu_0)^{\beta} $ \citep{Fruchter1990,Nice1990}, where $A$ is a normalization factor. We find a value of $\beta=-0.09 \pm 0.04$. Following previous examples from the literature (e.g. \citealt{Polzin2020}) we fit the frequency dependence of the position in orbital phase of the ingress, separately from the egress.
The ingress position is best fitted by a power-law $p_{1, \rm ing} (\nu) =p_{1, \rm ing, \nu_0} (\nu/\nu_0)^{\beta_{\rm ing}}$ with $\beta_{\rm ing}=-0.06 \pm 0.05$ while the egress position with a power-law $p_{1, \rm eg} (\nu) =p_{1, \rm eg, \nu_0} (\nu/\nu_0)^{\beta_{\rm eg}}$ with $\beta_{\rm eg}=-0.13 \pm 0.03$. Like in other cases of eclipsing pulsars \citep{Polzin2020, Wang2021}, the frequency dependence is steeper during the egress. 

These values are among the shallowest seen in eclipsing pulsars as shown in Table \ref{tab:eclipse_properties} but are still compatible with PSR J1720$-$0533 \citep{Wang2021}.


\section{Eclipse variability} \label{sec:eclipse_variability}

\begin{figure*}
\centering
	\includegraphics[width=0.49\textwidth]{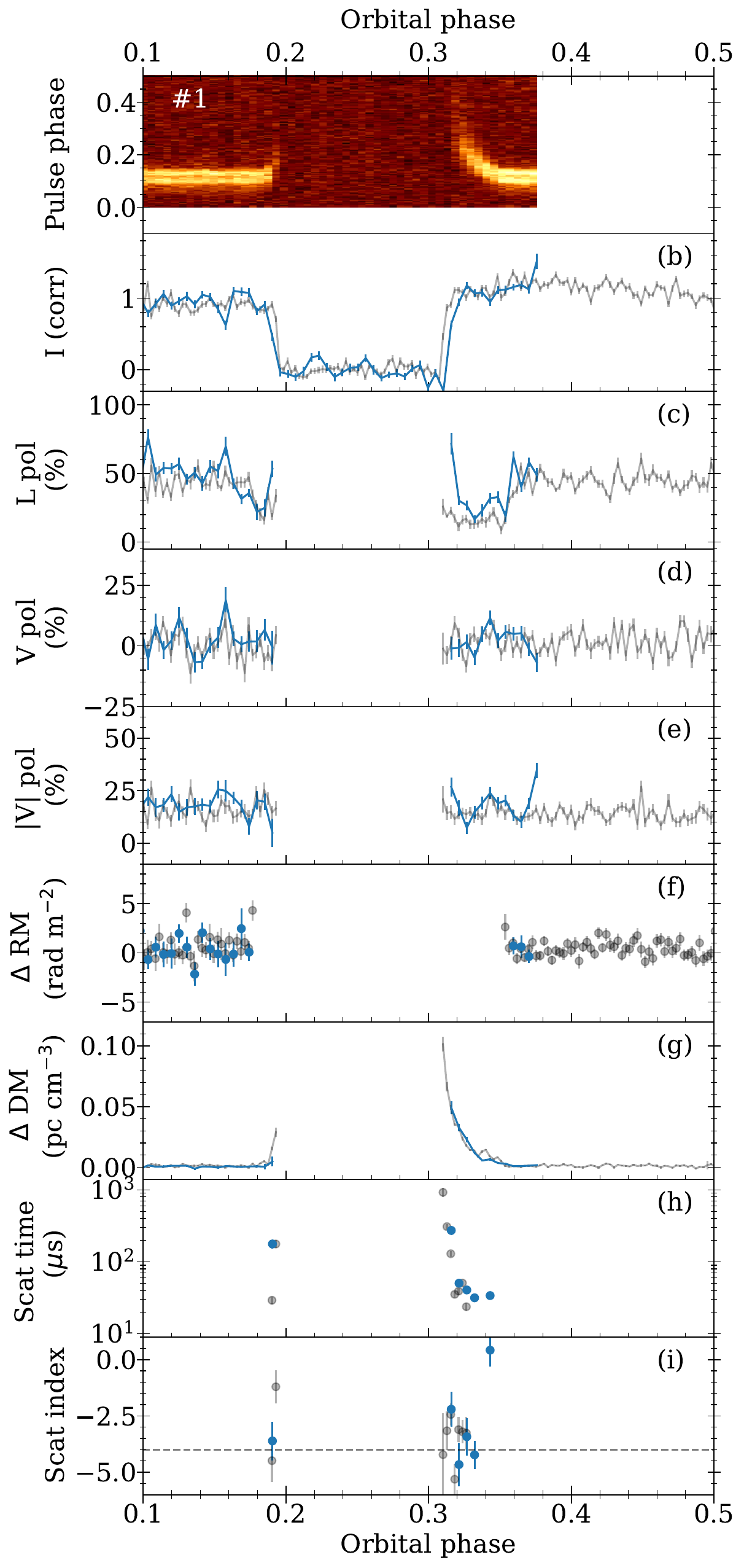}
 \,
	\includegraphics[width=0.49\textwidth]{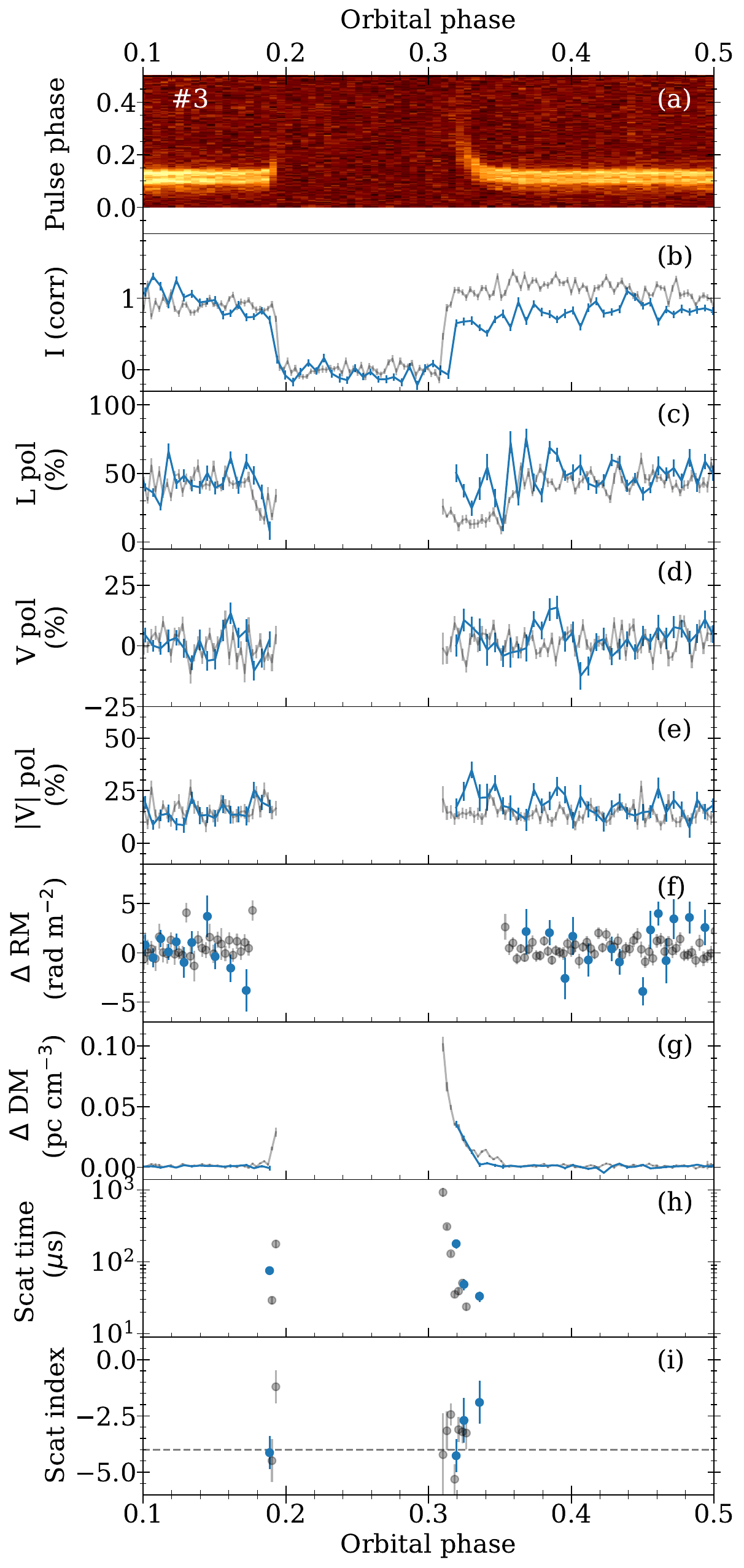}
    \caption{Same as Fig. \ref{fig:26U2_full} but for eclipse \#1 (left) and \#3 (right). For comparison we show in gray the measurements already reported in Fig. \ref{fig:26U2_full} for the case of eclipse \#2.}
  	\label{fig:26U13_full}
\end{figure*}

\begin{figure*}
\centering
	\includegraphics[width=0.49\textwidth]{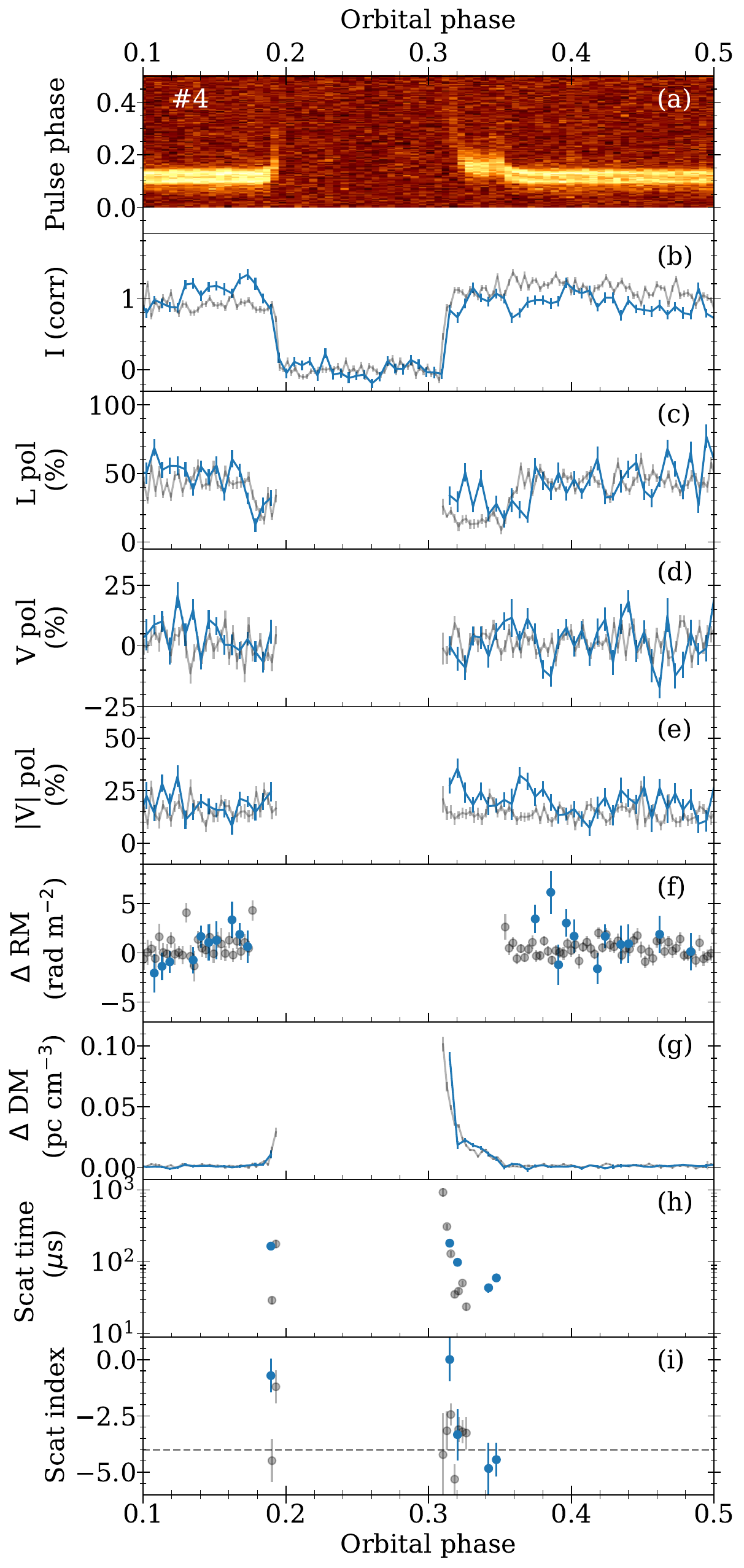}
 \,
	\includegraphics[width=0.49\textwidth]{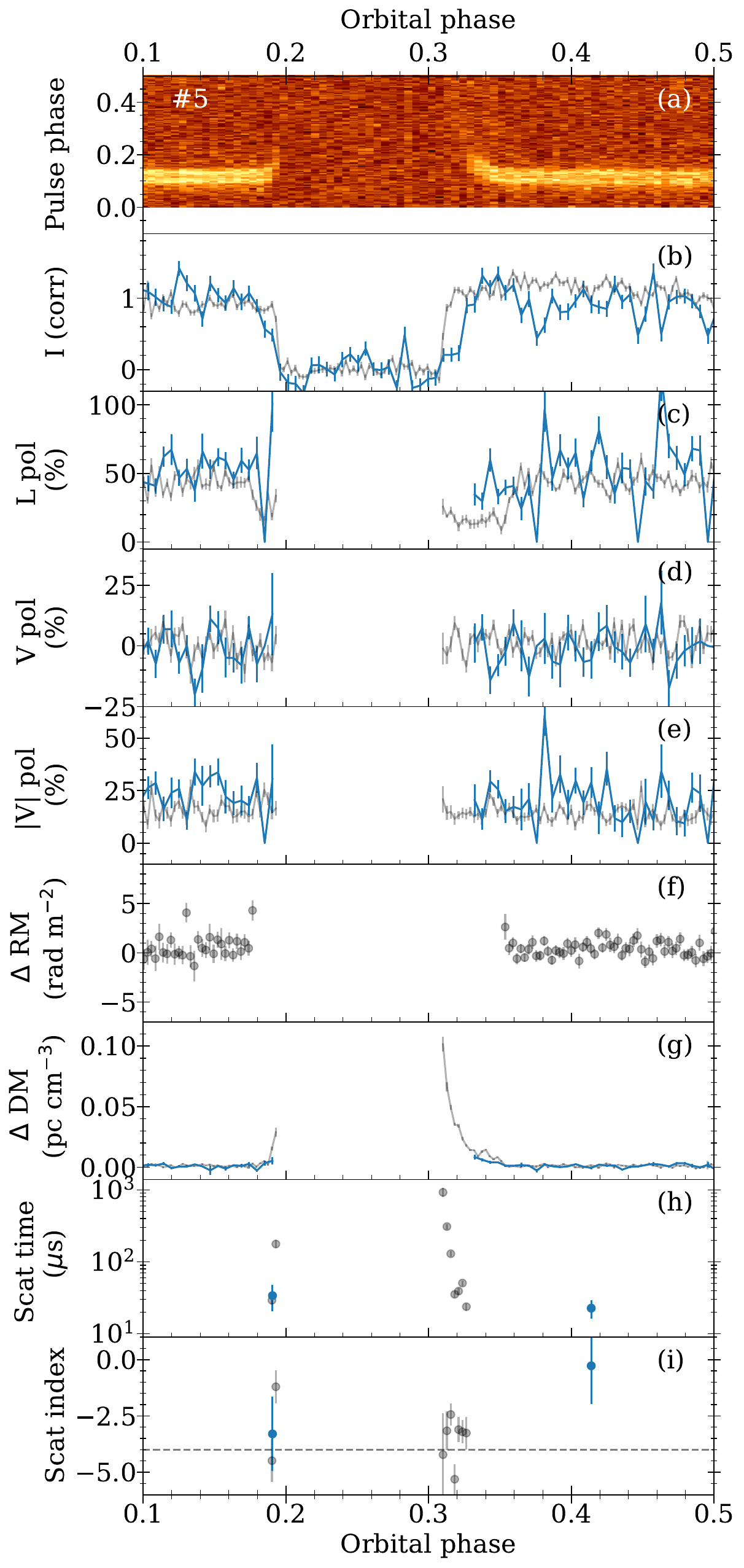}
    \caption{As for Figure \ref{fig:26U13_full} for the cases of the eclipses \#4 (left column) and \#5 (right column) of the 17 h observation of 47 Tuc O. }
  	\label{fig:26U45_full}
\end{figure*}

\begin{figure}
\centering
	\includegraphics[width=0.5\textwidth]{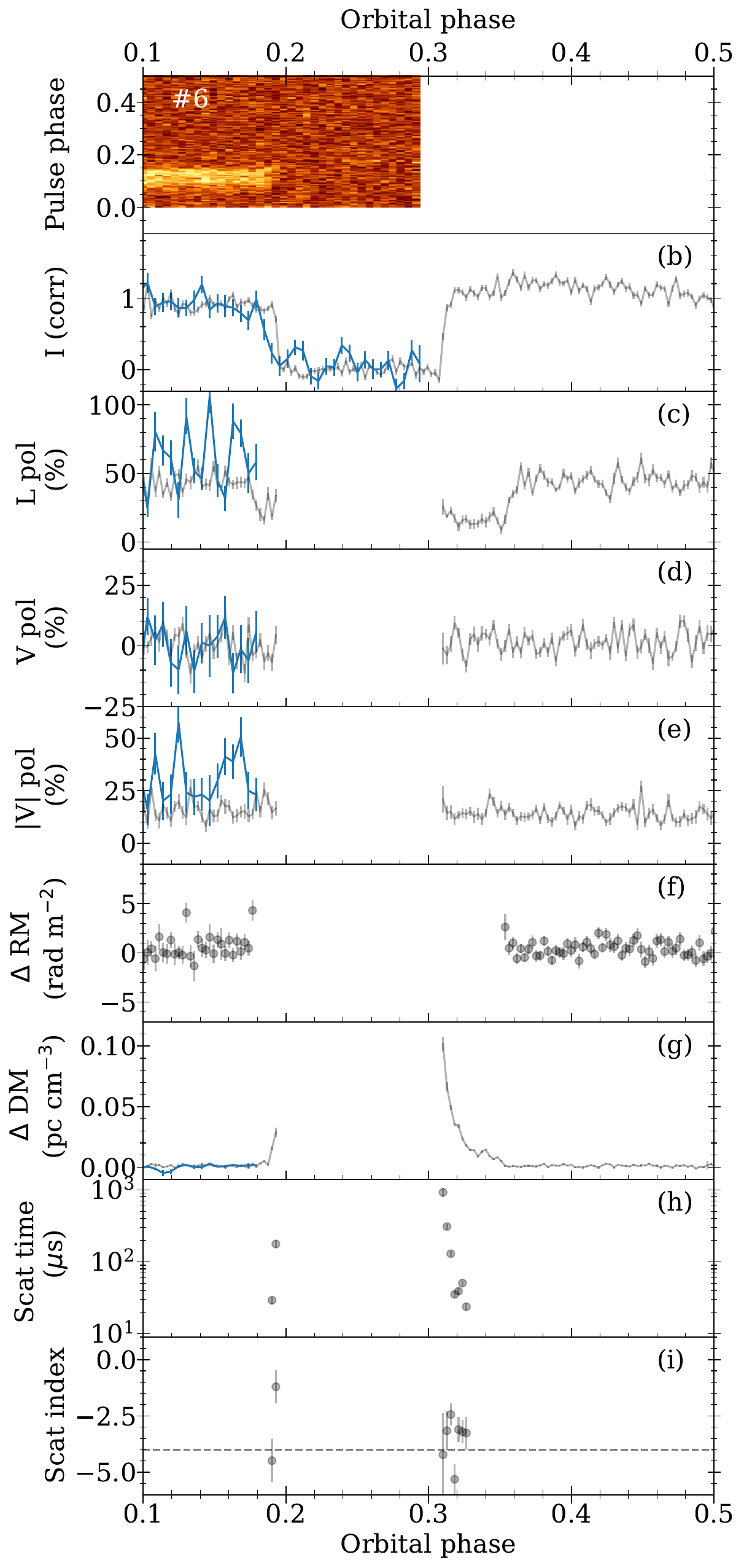}

    \caption{As for Figure \ref{fig:26U13_full} for the case of eclipse \#6 of the 17 h observation of 47 Tuc O.}
  	\label{fig:26U6_full}
\end{figure}

In Figures \ref{fig:26U13_full}, \ref{fig:26U45_full} and \ref{fig:26U6_full} we show the total flux, polarization percentages, RM and DM variations, scattering times and scattering indices for all the other eclipses visible during the observation. The data shown in these plots are reported in tables \ref{tab:data_26U1}, \ref{tab:data_26U3}, \ref{tab:data_26U4}, \ref{tab:data_26U5} and \ref{tab:data_26U6}. Due to the variable scintillation caused by the ISM during the day, the other eclipses do not appear as bright as eclipse \#2 and the total flux density around the eclipses varies. For this reason, we had to use 64s sub-integrations and the error bars are larger. The data for the RM, which is plotted only if the linearly polarized signal is strong enough, is sparser. 
The eclipse \#4 shows an extra delay during the egress transition and will be discussed in detail in Section \ref{sec:eclipse_4}.

\begin{table}
\centering 
\caption{Values of the start, end and total duration for the different eclipses. The start and end are defined as the best-fitting value of the parameter $p_1$ in the Fermi-Dirac function (eq. \ref{eq:fermi-dirac}). The length of the eclipse is the difference between these two values. For eclipse \#6 we don't have values for the egress since the observation ended while the pulsar was eclipsed.}
\label{tab:eclipse_lengths}
\centering
\renewcommand{\arraystretch}{1.0}
\vskip 0.1cm
\begin{tabular}{c|c|c|c|}
\hline
Eclipse \#   & ingress $p_1$  &  egress $p_1$ & Eclipse length \\
\hline
\#1 & 0.190(2) & 0.3171(8) & 0.127(3)\\
\hline
\#2 & 0.1935(9) & 0.3119(4) & 0.118(1) \\
\hline
\#3 & 0.188(1) & 0.317(2) & 0.129(3)\\
\hline
\#4 & 0.191(2) & 0.314(2) & 0.122(3)\\
\hline
\#5 & 0.188(2) & 0.322(3) & 0.135(5)\\
\hline
\#6 & 0.184(4) & - & - \\
\hline
\end{tabular}
\end{table}

\begin{figure}
\centering
	\includegraphics[width=0.5\textwidth]{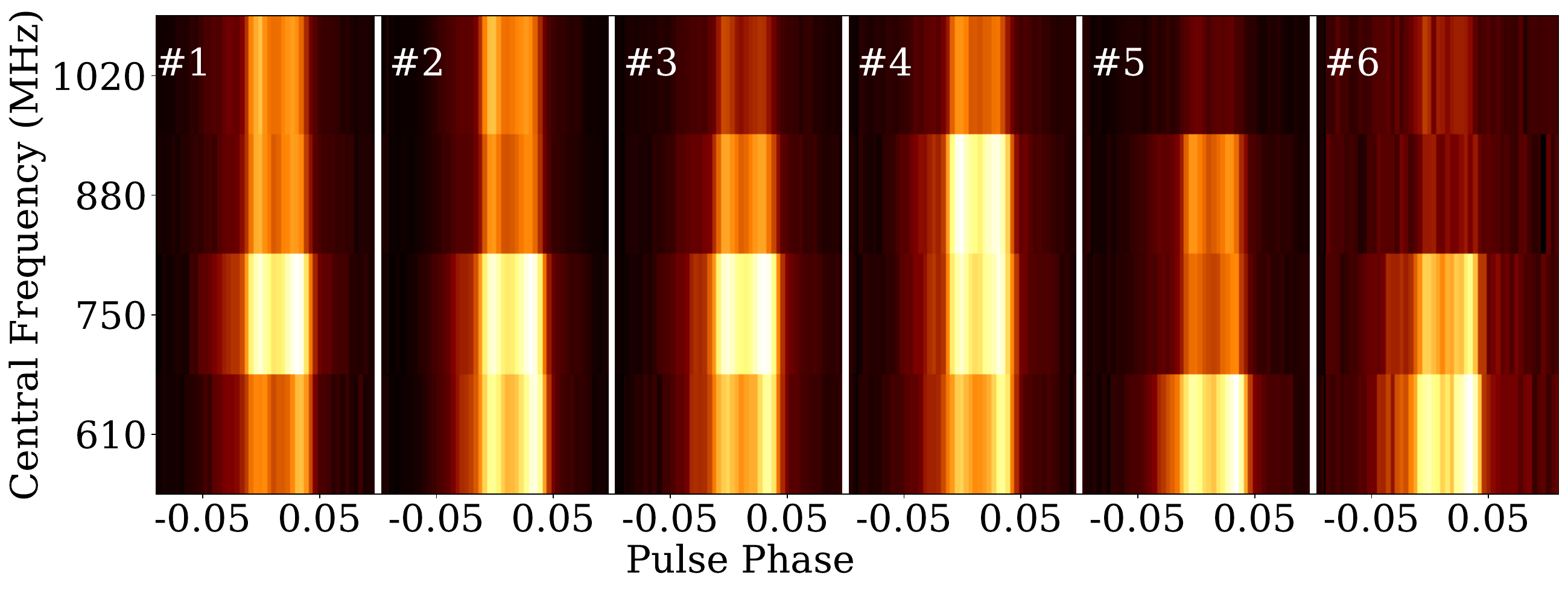}
    \caption{Plot showing the flux of 47 Tuc O divided into four frequency channels as a function of pulse phase across the different eclipses. The flux in each eclipse is normalized independently of the others with white representing the brightest frequency channel in each eclipse. The difference between the eclipses are caused by the scintillation due to the ISM. Eclipse \#5 and \#6 are significantly brighter in the lower frequency channel. }
  	\label{fig:frequency_evolution}
\end{figure}

The values of the ingress, egress and eclipse duration are reported in table \ref{tab:eclipse_lengths} using the definitions given in \S \ref{sec:eclipse_duration}. 
The lengths of the eclipses seem to vary significantly with every eclipse alternating between a longer and a shorter eclipse. 
The eclipse that lasts the longest is \#5. 
In this eclipse, due to scintillation, the top part of the frequency band has almost no signal. The contribution of each channel to the total flux during all the eclipses can be seen in Fig. \ref{fig:frequency_evolution}. In the case of eclipse \#5 and \#6 most of the signal comes from the lowest frequency channel. However, the frequency dependence of the eclipse cannot be the only explanation since the duration of eclipse \#2 at a comparable frequency band is only $0.120\pm 0.002$. The explanation is possibly found in varying conditions of the eclipsing medium.

A similar situation was detected in the case of PSR J1544+4937 \citep{Kumari2023} where the frequency cut-off of the eclipse is seen to vary over consecutive eclipses due to changes in the eclipsing medium. 
The alternating length of the eclipses might hint to a periodicity in the variability of the eclipsing medium.

\begin{figure}
\centering
	\includegraphics[width=0.49\textwidth]{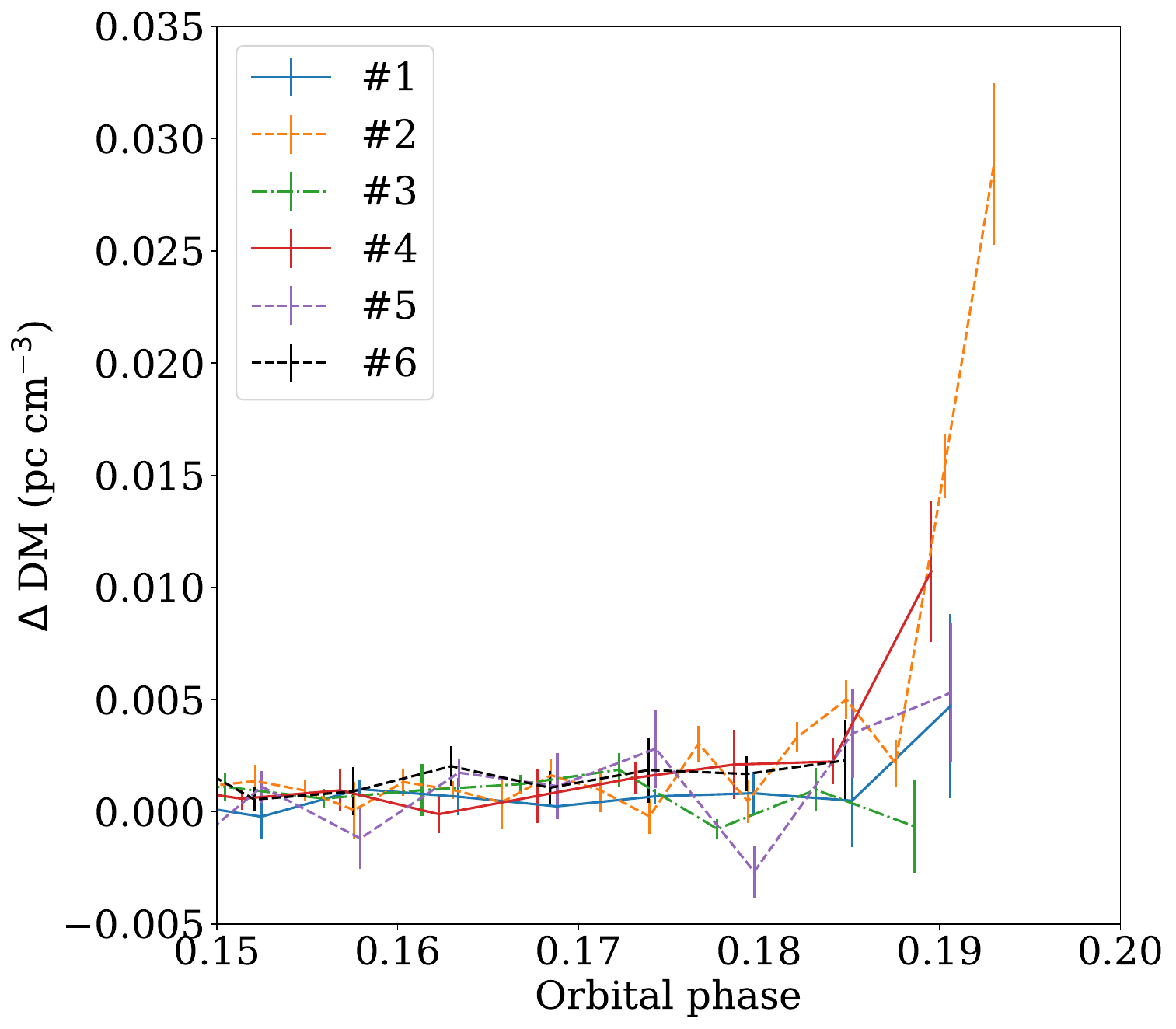}
	\includegraphics[width=0.49\textwidth]{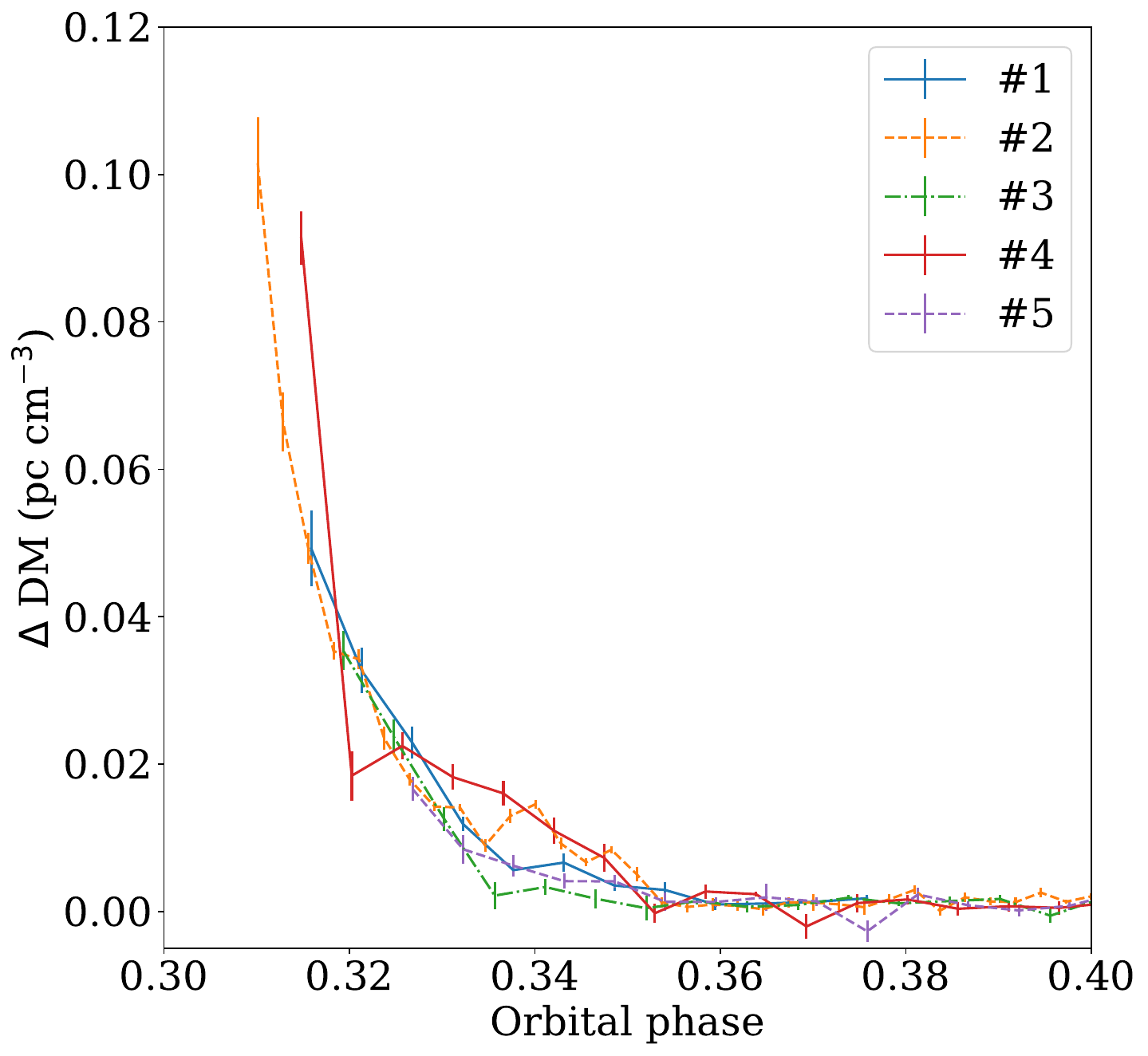}
    \caption{Variability of the DM excess in the ingresses (top panel) and egresses (bottom panels) of the 6 monitored eclipses. Both panels show the DM excess from the average value measured in {\protect \cite{Abbate2023}} as a function of the orbital phase. 
    Eclipse \#6 is not shown in the bottom panel since the egress was not visible in the observation. }
  	\label{fig:DM_12345}
\end{figure}

The DM excess shows significant eclipse-to-eclipse variations, during both the ingress and egress stage, as displayed in Figure \ref{fig:DM_12345}. The top and bottom panels shows the DM variations during the ingress and the egress respectively. At similar orbital phases, differences of more than 0.01 pc cm$^{-3}$ are visible in consecutive orbits. The eclipses \#2 and \#4 show increased DM with respect to the other orbits during the ingress and at the orbital phase of 0.34 during the egress. The eclipses with increased DM are also the ones that last the shortest possibly hinting at an anti-correlation between the duration of the eclipse and the DM at these orbital phases.
The large variability in consecutive orbits suggests that the gas in the eclipsing region has a variability timescale that is shorter than one orbital period.

We can use this to estimate a lower limit on the velocity of the fluctuations. We can only see fluctuations in the range of orbital phase between 0.185-0.19 in the ingress and 0.31-0.35 is the egress. Since the range in the egress is larger, it will give more stringent constraints on the velocity. To convert the range of orbital phases to a linear size, we can use the equation $R_{\rm eg}=2\pi \frac{M_p}{M_c} \frac{x}{\sin(i)} f_{\rm eg}$ where $f_{\rm eg}\sim 0.04$ is the range of orbital phases where fluctuations are seen in the egress, $M_p$ is the pulsar mass, $M_c$ is the companion mass, $x$ is the projected semi-major axis of the pulsar orbit and $i$ is the inclination angle of the system. From the timing of the pulsar \citep{Freire2017} we know that $x=0.0451533(3)$ lt-s and the mass function is 0.0000053461(1) M$_{\odot}$ while the masses and inclination angle are undetermined. Considering a range of possible pulsar masses between 1.1 M$_{\odot}$ and 2.1 M$_{\odot}$ (roughly encompassing the values of the pulsar masses which have been accurately measured so far), and inclination angles between 60 $\deg$ and 90 $\deg$ (thus spanning the most probable values for $i$) we can establish a range of possible values for $R_{\rm eg}$. The highest values of $R_{\rm eg}$ come from the case of a massive pulsar with a low orbital inclination. That gives $R_{\rm eg}\sim 0.35$ R$_{\odot}$. In contrast, the lowest values of $R_{\rm eg}$ come from a light pulsar with a high inclination. That gives $R_{\rm eg}\sim 0.28$ R$_{\odot}$.

In order for the DM fluctuations to traverse this region within one orbital period, the velocity must be larger than 17-20 km s$^{-1}$.


The scattering trends at both ingress and egress behaves similarly between the eclipses. In particular, the scattering time is very consistent between the first three eclipses, where it starts to rise at the same phase and reaches similar upper values. The scattering index values are also consistent, with a few exceptions, from eclipse to eclipse. The differences seen in some orbits are addressed in \S \ref{sec:scattering_index}.


\subsection{The case of eclipse \#4} \label{sec:eclipse_4}

The fourth eclipse manifests a unique feature: an extra delay in phase of the pulse profile during the egress stage, centred at the orbital phase 0.345 (see the panel (a) of the left column in Figure \ref{fig:26U45_full}). The occurrence of the delay is accompanied by an increase of the scattering time. The fourth eclipse also shows a delay in the recovery of the linear polarization percentage by about 4 64s sub-integrations (i.e. 0.022 orbital phases) with respect to the other monitored eclipses. This is coupled with a higher $\Delta$DM at later orbital phases and a significantly larger amount of scatter in the values for the $\Delta$RM right after the recovery, in comparison to other orbital phases. All these features can be observed by inspecting the left column of Figure \ref{fig:26U45_full}.

The lower panel of Figure \ref{fig:DM_12345} shows that the DM rise in these orbital phases was not present during the orbit immediately preceding or following. Interestingly, this panel also shows that the DM values reached during this event are similar to those seen in the eclipse \#2. Despite the similar values of DM, eclipse \#4 shows significantly stronger scattering. At those orbital phases scattering was undetected in eclipse \#2 meaning that it was below the threshold of 22.5 $\mu$s while in eclipse \#4 the measured scattering times are $44\pm 6$ and $60 \pm 7$ $\mu$s. 
In the simplest case of gaussian and isotropic thin-screen scattering \citep{Lang1971}, the scattering time is linearly related to the variance of the electron density fluctuations. This suggests that, despite a similar amount of total intervening gas, the variance of the fluctuations within a single sub-integration has increased at least by a factor of two.


The stronger electron density fluctuations also provide a natural explanation for the delay in the recovery of the linear polarization percentage as well as the increased scatter in the values of the $\Delta$RM. Stronger fluctuations of electron density within the same sub-integration can cause the depolarization to happen at a lower level of DM.
The $\Delta$DM in the sub-integrations where the linear polarization drops to 0 is $(1.1\pm 1.2)\times 10^{-3}$ pc cm$^{-3}$, much lower than the value measured in the eclipse \#2 of $\sim 6\times 10^{-3}$ pc cm$^{-3}$. Alternatively, the depolarization and the RM scatter could also be explained by a stronger magnetic field. Given that the excess DM is compatible with 0, we can only derive lower limits for the necessary magnetic field. If we assume that the depolarization is only caused by rapid fluctuations of direction of the magnetic field and following the reasoning already presented in \S\ref{sec:polarization}, we derive a 1$\sigma$ lower limit of the magnetic field of 2.5 mG and a 2$\sigma$ lower limit of 1.7 mG. These values are compatible with the measurements of eclipse \#2. However, if the magnetic field originates entirely from the companion it is unlikely to change in intensity over a small number of orbits.

We can use the measurements of DM during this event to determine the range of the variations in electron density over consecutive eclipses. 
The DM excess in the cloud of material responsible for the event during eclipse \#4 is 0.01 pc cm$^{-3}$ in the densest part compared to the previous and successive eclipse. We can measure the linear size of the projection along the orbital plane of the cloud $R_c$, by observing the fraction of the orbit influenced by it, $f_{\rm orb,c}\sim 0.03,$ and the equation $R_c=f_{\rm orb,c} 2\pi\frac{M_p}{M_c} \frac{x}{\sin i}$. The range of possible values of $R_c$ results (0.21-0.26) R$_{\odot}$. 
The electron density within the cloud can be determined by combining the value of the DM excess and the size of the cloud region $R_c,$ with the formula $n_e\sim \Delta {\rm DM}/ R_c$. If we assume that the cloud is roughly spherical and centered on the orbital plane, the excess DM and $R_c$ imply an electron number density in the range (1.7-2)$\times 10^6$ cm$^{-3}$. The number density of the surrounding gas measured at these orbital phases from the observation of the eclipses happening before and after is $\sim 1.3\times 10^5$ cm$^{-3}$, about a dozen times less than during the event. Finally, the total mass of the cloud under these assumptions is in the range 3-5$\times 10^{-20}$ M$_{\odot}$. Given the various assumptions, this value only represents a lower limit on the mass. 

\section{Discussion} \label{sec:discussion}

\subsection{Magnetic field}

The eclipses of 47 Tuc O show clear signs that the eclipsing material is permeated by a magnetic field. The linear polarization percentage drops to very low values as soon as the DM starts to increase due to the eclipsing material. Small, but measurable RM variations are detected in eclipse \#2 which, once coupled with the observed DM variations, can be explained by a magnetic field of $\sim 2$ mG. A magnetic field of the same order of magnitude appears to be present at the location of the enhanced scattering event visible during eclipse \#4. 

Generally, the magnetic field constraints in the companion of the eclipsing pulsars are determined by equating the pulsar wind pressure with the magnetic pressure of the companion. These calculations return values between 10-40 G \citep{Thompson1994,Wang2021} and refer to the magnetic field at the surface where the pressures are equal. Measurements of the magnetic fields in the eclipsing material derived from the assumption that the eclipses are caused by cyclotron-synchrotron absorption return similar values of 10-20 G \citep{Kudale2020,Kansabanik2021}. On the other hand, observations of RM changes and plasma lensing have determined the magnetic field in the eclipsing material to be closer to $\sim 1$-100 mG \citep{Crowter2020,Wang2023}. Recently, Faraday conversion has been measured in PSR B1744$-$24A implying a magnetic field stronger than 10 G at superior conjunction \citep{Li2023}. For the same pulsar, at a different orbital phase, \cite{You2018} and \cite{Li2023} observed linear depolarization and RM variations that suggested a magnetic field of $\sim 10$ mG.

For 47 Tuc O using both linear depolarization and RM variations we measured a magnetic field of the order of 2 mG and are in line with those of \cite{Crowter2020} who also used the RM variations to determine the magnetic field but smaller than the results of \cite{Wang2023}. The relatively low magnetic field suggests that cyclotron-synchrotron absorption should not be an effective mechanism for the eclipse.

In the assumption that the measured magnetic field belongs to the companion, we can estimate the strength at its surface. If the magnetic field has a dipolar geometry, we can scale the values to the surface using the equation $B_S = B_e (d/R_S)^3$, where $B_e$ is the magnetic field in the eclipsing material, $d$ is the distance of the eclipse to the companion and $R_S$ is the size of the companion. Taking the second orbit as an example, the orbital phase where we see the increase of RM is 0.177 during the ingress and 0.354 during the egress. The linear distance travelled by the companion during this time is $R=2\pi \frac{M_p}{M_c} \frac{x}{\sin(i)} f_{\rm RM}$ where $f_{\rm RM}$ is the difference in orbital phases between superior conjunction and the point where the increased RM is detected. The linear distance between the pulsar and the companion projected on the plane of the sky at this moment of the orbit can be expressed as: $d=\sqrt{R^2+ (a\cos(i)\cos(2\pi f_{\rm RM}))^2}$, where $a=\frac{M_p}{M_c} \frac{x}{\sin(i)}$ is the deprojected distance between the pulsar and the companion. 

In order to determine the magnetic field at the surface of the companion we need to estimate its size. The companion is constantly losing mass and therefore does not have a stable surface we can refer to. As a reference we can take the Roche lobe that defines the radius inside which the material is gravitationally bound to the companion. The size of the Roche lobe can be approximated by \citep{Eggleton1983}:
\begin{equation}
    R_S \simeq R_L \sim a \frac{0.49\, q^{2/3}}{0.6\, q^{2/3} + \ln(1+q^{1/3})},
\end{equation}
where $q=M_p/M_c$. Given the range of possible pulsar masses between 1.1 M$_{\odot}$ and 2 M$_{\odot}$ and inclination between 60 $\deg$ and 90 $\deg$, the estimated magnetic field at the Roche lobe becomes 100-210 mG.

This estimate is still only a small percentage of the magnetic field estimates at the surface of the companion in other eclipsing pulsar systems. However, the RM is only sensitive to the component of the magnetic field parallel to the line of sight, so, depending on the orientation of magnetic field lines in the eclipsing material, we might be sensitive to only a fraction of the total field strength. Furthermore, dipolar geometry might not hold in the highly perturbed and variable environment surrounding the companion star. Therefore, it is reasonable to assume that the measured value of the magnetic field is overall compatible with the previous literature.

\subsection{Mass loss}
\label{subs:mass loss}

The gas responsible for the eclipse is located outside the Roche lobe of the companion \citep{Freire2003} so it is not gravitationally bound anymore to the companion and is being ejected. From the measurements of the DM excess, we can estimate the mass loss rate due to the escaping material. 

First we must estimate the density of the gas on the border of the eclipse. The observed DM excess in the egress at the orbital phase of 0.310, where the total intensity flux drops, is $0.101\pm0.006$ pc cm$^{-3}$.
The de-projected linear size of the region where the pulsar is eclipsed, $R_e$, is derived from the range in orbital period of the eclipse as $R_e=2\pi \frac{M_p}{M_c} \frac{x}{\sin(i)} T_e/2$, where $T_e\sim 0.12$ is the full duration of the eclipse in orbital phase. The range of $R_e$ using the most probable values of masses and inclination angles is (0.42-0.52) R$_{\odot}$. In the approximation of a spherical outflow, we can consider the eclipsed region as a spherical shell of constant density. This approximation does not completely describe the system as we detect differences in the outflow during the ingress and the egress but can regardless return a good estimate of the expected average density. 
The average electron number density in the spherical shell is determined by using the formula $n_e\sim \Delta {\rm DM}/ R_e$ in the range (8-10)$\times 10^{6}$ cm$^{-3}$.

The mass loss due to this material can be estimated by assuming that the pulsar wind energy is converted to kinetic energy of the eclipsing material \citep{Thompson1994}. The pulsar wind energy density $U_E$ at the distance of the companion is given by $U_E=\dot E /(4\pi c a^2)$, where $c$ is the speed of light, $a$ is the distance from the pulsar to the companion and $\dot E$ is the spin down luminosity of the pulsar given by $\dot E = 4\pi^2 I \dot P P^{-3}$, where $I$ is the moment of inertia of the pulsar assumed to be $10^{45}$ g cm$^{2}$, $P$ is the rotational period and $\dot P$ is the intrinsic spin-down. If this energy density is converted to kinetic energy with an efficiency $\epsilon$, the outflow velocity $V_w$ is given by $V_w \sim (\frac{2\epsilon U_E}{n_e m_p} )^{1/2}$ \citep[eq. 7]{Yan2021}, where $n_e$ is the density of the outflowing gas and $m_p$ is the proton mass. With this velocity, the mass-loss rate can be quantified as the mass density in the shell multiplied by the surface area of the shell and the outflow velocity, ${\dot M}_c \simeq \pi R_e^2 m_p n_e V_w$ \citep{Thompson1994}.

The intrinsic period derivative $\dot P$ cannot be taken directly from the measured $\dot P$ of the pulsar because the acceleration due to the globular cluster can influence this quantity \citep{Abbate2018}. A measure of the orbital period derivative would allow us to disentangle the globular cluster contribution, but such a measure was not possible for this pulsar \citep{Freire2017}. Therefore we adopt, as a reference value, the average of the intrinsic spin-downs of the other binaries in the cluster with a measured value of the orbital period derivative \citep{Freire2017}. We obtain $\dot P \sim 1\times 10^{-20}$ s s$^{-1}$. Assuming, as usual, that the moment of inertia of the pulsar is $I\sim10^{45}$g cm$^2$, the spin-down energy results $\dot E\sim 2 \times 10^{34}$ erg s$^{-1}$. Spanning a range of pulsar masses and orbital inclinations as above, the distance to the companion $a$ is found to be in the range 1.1-1.4 R$_{\odot}$. Introducing an average density of the gas as calculated above (i.e. in the range of (8-10)$\times 10^{6}$ cm$^{-3}$), the ejection velocity of the wind results in the range (0.9-1.0)$\times 10^7$ m s$^{-1}$. Therefore, the mass loss rate is eventually found to be in the range (7.6-8.4)$\times 10^{-13}$ M$_{\odot}$ yr$^{-1}$. Given that the companion mass is between 0.02 and 0.03 M$_{\odot}$, it would take $\sim 30$ Gyr to completely evaporate the companion. This mass loss is similar to the value of $\sim 10^{-12}$ M$_{\odot}$ yr$^{-1}$ that has been measured for other eclipsing pulsars like J1810+1744 \citep{Polzin2018}, J2051-0827 \citep{Polzin2019}, and B1957+20 \citep{Polzin2020}. We note that if the system is not almost edge-on (i.e. the inclination is significantly lower than 90$^{\circ})$ and the mass ejection is concentrated on the orbital plane, the line of sight will not pass through the densest part. In this case our mass loss estimate would only capture a small fraction of the total one. Furthermore, the approximation of a spherical outflow can lead to an error of the estimate.

\subsection{Scattering index}\label{sec:scattering_index}

The scattering index during the eclipses is, with some exceptions, within the range of $-5$ to $-2.5$. This roughly corresponds to the similar range observed for other eclipsing pulsars like PSR B1957+20 \cite{Bai2022} and PSR J1720$-$0533 \cite{Wang2021}.

As anticipated in \S\ref{intro}, the expected value for the scattering index goes from $-4$ for an isotropic Gaussian spectrum \citep{Cronyn1970,Lang1971} to $-4.4$ for a Kolmogorov spectrum \citep{Lee1976,Rickett1977}. In case of homogeneous and isotropic turbulence this value cannot get higher than $-4$ \citep{Romani1986}. 
Deviations from these values can be expected if the scattering is happening in a finite region of comparable size to the scale at which the phase difference of the scattered rays with respect to the straight-line path is 1 radian \citep{Cordes2001}. In this situation, the scattering region cannot be larger than the physical size of the screen and the frequency dependency becomes flatter. This in turn results in a break in the frequency dependency of the scattering tail that can appear as a shallower index in the fit. Alternatively, scattering from an anisotropic screen could also lead to a shallower scattering index \citep{Stinebring2001,Geyer2017}. Additionally, the scattering index can also be affected by scintillation which can boost the flux in some parts of the frequency band while reducing it in other parts.  
These effects might explain the observed range of scattering indices at the rims of the monitored eclipses.

\section{Conclusions} \label{sec:conclusions}

\begin{table*}
\centering 
\caption{Comparison between the properties of J0024$-$7204O and other eclipsing pulsars. The type of each pulsar, whether black widow (BW) or redabck (RB) is listed next to the name. We report, where possible, the duration of the eclipse in percent of the orbit, the maximum DM excess measured, the frequency dependence index, the proposed mechanism for the eclipse and the estimated mass loss rate. The number in parenthesis refers to the frequency in MHz at which the measurement was done. References: $^a$ This work, $^b$ \citep{Archibald2009}
$^c$ \citep{Kudale2020}, $^d$ \citep{Deneva2016}, $^e$ \citep{Bhattacharyya2013}, $^f$ \citep{Kumari2023}
,$^g$ \citep{Vleeschower2024}, $^h$ \citep{Wang2021}, $^i$ \citep{Nice1990}, $^j$ \citep{Li2023}, $^k$ \citep{Polzin2018}, $^l$ \citep{Polzin2020}, $^m$ \citep{Yan2021}, $^n$ \citep{Polzin2019}, $^o$ \citep{Broderick2016}, $^p$ \citep{Crowter2020}. 
}
\label{tab:eclipse_properties}
\centering
\renewcommand{\arraystretch}{1.0}
\vskip 0.1cm
\begin{tabular}{c|c|c|c|c|c}
\hline
Name	& \multicolumn{1}{c}{Eclipse duration} & \multicolumn{1}{c}{DM excess}&	\multicolumn{1}{c}{Frequency dependence index} & Proposed mechanism & \multicolumn{1}{c}{Mass loss} \\
 & \multicolumn{1}{c}{(percent of the orbit)} & \multicolumn{1}{c}{ (pc cm$^{-3}$)} & \multicolumn{1}{c}{$\beta$} &  & \multicolumn{1}{c}{(M$_{\odot}$ yr$^{-1}$)}\\
\hline
J0024-7204O (BW)	& 12(812) $^a$	& 0.1(812) $^a$&	$-0.09\pm 0.04$ $^a$& scattering$^a$& $(7.6-8.4) \times 10^{-13}$ $^a$\\
\hline
J1023+0038	(RB)	& 40(685) $^b$ &0.15(700) $^b$ &$-$0.41 $^c$ & - &  -\\		
\hline
J1048+2339	(RB)	& 57(327) $^d$ &0.008(327) $^d$ & - & - & -\\	
\hline
J1227$-$4853 (RB)& 64(607) $^c$ & 0.079(607) $^c$ &$-$0.44 $^c$& Cyclotron damping $^c$&  - \\
\hline	
J1544+4937 (BW) & 13(322) $^e$ & 0.027(607)  $^e$& - & -	& (2-25) $\times 10^{-14}$ $^f$ \\
\hline
J1701-3006B (BW) & 25(714) $^g$ & 0.6 (1123) $^g$ & $-$0.35 $^g$&  -&  -\\
\hline
J1720$-$0533 (BW)	& 13(1250)	$^h$ & 0.6(1250) $^h$ &$-0.14\pm 0.07$ $^h$&	poss. scattering $^h$ & -\\
\hline 
B1744$-$24A (RB)	& 50(1000) $^i$	& 0.4(1499) $^j$	&$-0.63 \pm 0.18$ $^i$  & - & - \\
\hline
J1810+1744 (BW)	& 13(149) $^k$& 0.01(345) $^k$ &	$-0.22\pm 0.02$ $^k$ & - &$10^{-12}$ $^k$\\
\hline 
J1816+4510 (RB)	& 11(150) $^l$	&0.01(149) $^l$&	$-0.49 \pm 0.06$ $^l$&	scattering $^l$ &$2\times 10^{-13}$ $^l$\\
\hline
J1853$-$ 0842A (BW)	& - &0.03(1250) $^m$& - & - &  $3 \times 10 ^{13}$ $^m$\\
\hline
B1957+20  (BW)	& 18(121) $^l$& 0.01(149) $^l$ &$-0.18\pm 0.04$ $^l$	& \multicolumn{1}{c}{scattering$^m$}& $10^{-12}$ $^l$\\
& &  & 	& \multicolumn{1}{c}{ cyclotron dampening$^m$}& \\
\hline
J2051$-$0827 (BW)	& 14(149) $^l$ &0.1(2364) $^n$& $-0.41 \pm 0.04 $ $^l$ & \multicolumn{1}{c}{scattering$^m$} & $10^{-12}$ $^n$\\
& &  & 	& \multicolumn{1}{c}{ cyclotron dampening$^m$}& \\
\hline
J2215+5135 (RB)	& 50(149) $^o$ & - & $-0.21 \pm 0.04$ $^l$ & - & - \\		
\hline
J2256$-$1024 (BW) & 5(820) $^p$ & 0.06 (820) $^p$ & - & - & - \\
\hline
\end{tabular}
\end{table*}

We observed the eclipsing black widow pulsar 47 Tuc O with the MeerKAT radio telescope at UHF band (544-1088 MHz) for 17 hours, covering 6 contiguous orbits (but only 5 full eclipses). Thanks to this observation we have been able to study in detail the properties of the eclipses and their variability over a succession of orbits. In particular, we have reported the total intensity, polarization parameters, excess RM, excess DM, scattering time, and scattering index for each of the monitored eclipses with 64s time resolution. Thanks to a boost in the signal caused by scintillation in the ISM, we were able to study the eclipse \#2 in greater detail with a time resolution of 32s.

The eclipses are characterized by increased values of DM and scattering in both the ingress and egress transition stages, which imply a higher gas density along the line of sight within the eclipsing region than in the surrounding orbital phases. The strong scattering, the speed at which it grows close to the edge of the eclipse and the reduction in flux point to scattering being the dominant mechanism that causes the eclipses, at least in proximity of the ingress and egress regions. A simultaneous continuum imaging observation of the pulsar was performed at MeerKAT and the results will be presented in a future publication.


In Table \ref{tab:eclipse_properties} we summarize the properties of the eclipses of J0023$-$7204O and compare them to the values for other eclipsing pulsars. The duration of the eclipse is similar to that of other black widow pulsars. The maximum measured DM excess is comparable to that of the other eclipsing pulsars that were observed in a similar frequency range. The frequency dependence of the eclipse is shallower than the other eclipsing pulsars but still compatible with PSR J1720$-$0533 \citep{Wang2021} that also shows potential signs of being dominated by scattering.

The linear polarization shows a very significant drop during both the ingress and egress transition stages. During eclipse \#2, the drop occurs when the DM excess reaches $\sim 0.003$-0.005 pc cm$^{-3}$. This depolarization, seen in other eclipsing pulsars \citep{You2018, Polzin2019}, is caused by large fluctuations of the RM within a single 32s sub-integration. The eclipse also shows an increase of RM right before the drop of linear polarization, which indicates a magnetic field of order $\sim 2$ mG. This value is similar to other measurements of eclipsing pulsars made with the same method \citep{Crowter2020}. If we assume that the magnetic field is linked to the companion star we can calculate a value of 100-210 mG for the magnetic field near the surface. Given the wide range of assumptions we cannot exclude that it is overall compatible with the other values of the magnetic field reported for the surface of the companions of other eclipsing pulsars.

The absolute value of the circular polarization in eclipses \#1, \#2 and \#4 shows a small peak that could be explained in the context of induced circular polarization by multipath propagation through a magnetized scattering screen \citep{Beniamini2022}. 

Thanks to the unprecedented length of the observation for this kind of targets, it was possible to study the variability of the eclipses in several consecutive orbits. The eclipse length, DM excess, scattering and polarization percentages all show significant changes on an orbit-by-orbit basis. The eclipses lengths and DM excess show alternating patterns in consecutive orbits  hinting at a possible periodicity in properties of the eclipsing material. 
The DM shows variations of $\sim 0.01$ pc cm$^{-3}$ over consecutive orbits at the orbital phases at which the eclipses occur. This implies that the material ejected by the companion can traverse the eclipsing region in less than an orbit. The most significant event occurred during eclipse \#4, where an extra delay is seen in the pulse profile versus orbital phase. This event is coincident with an excess of DM, increase in scattering time of a factor of 2, delayed recovery of the linear polarization of $ 0.02$ orbital phases and increased scatter of RM. This event is most likely caused by a cloud of denser material in the out-flowing gas. This cloud contributes to $\sim 0.01$ pc cm$^{-3}$ to the total DM and is $\sim 12$ times denser than the surrounding gas. We suggest that the higher scattering time, delayed recovery of linear polarization and higher scatter of RM are caused by an increase of the variance of the electron density fluctuations in the cloud.

This study reveals the importance of monitoring the variability of successive orbits of an eclipsing pulsar. This is especially true in the case of scintillating sources like 47 Tuc O. Not only were we able to quantify electron density fluctuations on short timescales, but we were also able to study the details of an eclipse at a high state of scintillation. While it is not possible to predict the occurrence of a high scintillation state, observing consecutive orbits strongly increases the probability of it happening during the observation. Even in cases where scintillation is not as issue, monitoring the orbit-to-orbit variability can help constrain the properties of the eclipsing gas and of the under-lying mechanism. Therefore, this observing strategy can be applied to other eclipsing pulsars, especially to the more rapidly variable red-backs and other scintillating sources.


\section*{Acknowledgements}

The MeerKAT telescope is operated by the South African Radio Astronomy Observatory, which is a facility of the National Research Foundation, an agency of the Department of Science and Innovation. SARAO acknowledges the ongoing advice and calibration of GPS systems by the National Metrology Institute of South Africa (NMISA) and the time space reference systems department department of the Paris Observatory. PTUSE was developed with support from the Australian SKA Office and Swinburne University of Technology. MeerTime data is housed on the OzSTAR supercomputer at Swinburne University of Technology. The OzSTAR program receives funding in part from the Astronomy National Collaborative Research Infrastructure Strategy (NCRIS) allocation provided by the Australian Government. We thank the SARAO technical teams that developed and implemented the 4-beam steerable system used in this work. 
Part of this work has been funded using resources from the INAF Large Grant 2022 “GCjewels” (P.I. Andrea Possenti), approved with the Presidential Decree 30/2022. This work was also supported in part by the “Italian Ministry of Foreign Affairs and International
Cooperation”, grant number ZA23GR03. 
FA and AP acknowledge that part of the research activities described in this paper were carried out with the contribution of the NextGenerationEU funds within the National Recovery and Resilience Plan (PNRR), Mission 4 - Education and Research, Component 2 - From Research to Business (M4C2), Investment Line 3.1 - Strengthening and creation of Research Infrastructures, Project IR0000034 – “STILES -Strengthening the Italian Leadership in ELT and SKA”. 

\section*{Data Availability}

The data underlying this article will be shared upon reasonable request to the MeerTime and TRAPUM collaborations.



\bibliographystyle{mnras}
\bibliography{biblio} 



\appendix

\section{Data tables}\label{appendix:data}

The tables \ref{tab:data_26U1},\ref{tab:data_26U2},\ref{tab:data_26U3},\ref{tab:data_26U4},\ref{tab:data_26U5} and \ref{tab:data_26U6} contain the results of the analysis used to create Fig. \ref{fig:26U2_full},\ref{fig:26U13_full},\ref{fig:26U45_full} and \ref{fig:26U6_full}. 

\begin{table*}
\centering 
\caption{Table containing the orbital phase, corrected integrated flux, linear, circular and absolute circular polarization percentages, the RM and DM excesses, the scattering times and the scattering indices during eclipse \#1. The DM and RM excesses are determined in comparison with the values presented in {\protect\cite{Abbate2023}}. We only report the values in the orbital phase range 0.15-0.4 and exclude the sub-integrations where the signal is eclipsed. The excess RM is reported only if a clear detection was possible and the scattering parameters are reported only if the scattering time exceeds 22.5 $\mu$s, corresponding to 3 times the sampling time. All the errors are reported at 1$\sigma$.}
\label{tab:data_26U1}
\centering
\renewcommand{\arraystretch}{1.0}
\vskip 0.1cm
\begin{tabular}{c|c|c|c|c|c|c|c|c|c}
\hline
Eclipse \#   & \multicolumn{1}{c}{orbital} & I (corr)&  L pol percent & V pol percent & $|$V$|$ pol percent &\multicolumn{1}{c}{$\Delta$ RM} & \multicolumn{1}{c}{$\Delta$ DM} &\multicolumn{1}{c}{Scat. time} & Scat. index \\
&  \multicolumn{1}{c}{(phase)}  & & & & & \multicolumn{1}{c}{(rad m$^{-2}$)} & \multicolumn{1}{c}{(pc cm$^{-2}$)} & \multicolumn{1}{c}{($\mu$s)} & \\
\hline
1 (ing) & 0.152 & 0.85 ( 6 ) & 52 ( 5 ) & 3 ( 4 ) & 26 ( 4 ) & 0 ( 1 ) & $-$0.0002 ( 10 ) &   &   \\
\hline
& 0.158 & 0.62 ( 6 ) & 70 ( 7 ) & 19 ( 5 ) & 25 ( 5 ) & $-$1 ( 2 ) & 0.0010 ( 4 ) &   &   \\
\hline
& 0.163 & 1.10 ( 6 ) & 43 ( 4 ) & 3 ( 3 ) & 22 ( 3 ) & $-$0 ( 1 ) & 0.0007 ( 8 ) &   &   \\
\hline
& 0.169 & 1.08 ( 6 ) & 31 ( 4 ) & 1 ( 3 ) & 18 ( 3 ) & 2 ( 2 ) & 0.0002 ( 4 ) &   &   \\
\hline
& 0.174 & 1.07 ( 6 ) & 36 ( 3 ) & 2 ( 4 ) & 8 ( 4 ) & 0.1 ( 9 ) & 0.0007 ( 3 ) &   &   \\
\hline
& 0.180 & 0.82 ( 5 ) & 22 ( 6 ) & 2 ( 5 ) & 20 ( 4 ) &   & 0.001 ( 1 ) &   &   \\
\hline
& 0.185 & 0.91 ( 5 ) & 25 ( 6 ) & 7 ( 4 ) & 20 ( 4 ) &   & 0.001 ( 2 ) &   &   \\
\hline
& 0.191 & 0.46 ( 6 ) & 53 ( 6 ) & $-$1 ( 7 ) & 5 ( 7 ) &   & 0.005 ( 4 ) & 180 ( 20 ) & $-$3.6 ( 9 ) \\
\hline
\hline
1 (eg )& 0.316 & 0.64 ( 5 ) & 72 ( 8 ) & $-$1 ( 5 ) & 27 ( 5 ) &   & 0.049 ( 5 ) & 270 ( 4 ) & $-$2.2 ( 8 ) \\
\hline
& 0.321 & 0.94 ( 5 ) & 30 ( 3 ) & -1 ( 4 ) & 17 ( 4 ) &   & 0.032 ( 3 ) & 51 ( 6 ) & $-$5 ( 1 ) \\
\hline
& 0.327 & 1.18 ( 5 ) & 27 ( 3 ) & 2 ( 3 ) & 7 ( 3 ) &   & 0.023 ( 2 ) & 41 ( 5 ) & $-$3.4 ( 8 ) \\
\hline
& 0.332 & 1.07 ( 6 ) & 17 ( 3 ) & -5 ( 3 ) & 15 ( 3 ) &   & 0.0119 ( 9 ) & 32 ( 3 ) & $-$4.2 ( 6 ) \\
\hline
& 0.338 & 1.08 ( 6 ) & 23 ( 5 ) & 5 ( 3 ) & 19 ( 3 ) &   & 0.0056 ( 8 ) &   &   \\
\hline
& 0.343 & 0.95 ( 5 ) & 32 ( 4 ) & 11 ( 3 ) & 24 ( 3 ) &   & 0.007 ( 1 ) & 34 ( 4 ) & 0.4 ( 7 ) \\
\hline
& 0.349 & 1.11 ( 5 ) & 33 ( 4 ) & 2 ( 3 ) & 19 ( 3 ) &   & 0.0035 ( 7 ) &   &   \\
\hline
& 0.354 & 1.12 ( 5 ) & 19 ( 4 ) & 6 ( 3 ) & 20 ( 3 ) &   & 0.0030 ( 1 ) &   &   \\
\hline
& 0.359 & 1.15 ( 5 ) & 62 ( 4 ) & 5 ( 3 ) & 13 ( 3 ) & 0.7 ( 9 ) & 0.0009 ( 7 ) &   &   \\
\hline
& 0.365 & 1.18 ( 6 ) & 40 ( 4 ) & 5 ( 3 ) & 10 ( 3 ) & 1 ( 1 ) & 0.0011 ( 5 ) &   &   \\
\hline
& 0.370 & 1.12 ( 6 ) & 58 ( 3 ) & $-$1 ( 3 ) & 19 ( 3 ) & $-$0.4 ( 7 ) & 0.0013 ( 6 ) &   &   \\
\hline
& 0.376 & 1.51 ( 11 ) & 49 ( 5 ) & $-$7 ( 4 ) & 35 ( 4 ) &   & 0.0018 ( 4 ) &   &   \\
\hline
\end{tabular}
\end{table*}

\begin{table*}
\centering 
\caption{Same as Table \ref{tab:data_26U1} for the ingress of eclipse \#2 using 32s sub-integrations.}
\label{tab:data_26U2}
\centering
\renewcommand{\arraystretch}{1.0}
\vskip 0.1cm
\begin{tabular}{c|c|c|c|c|c|c|c|c|c}
\hline
Eclipse \#   & \multicolumn{1}{c}{orbital} & I (corr)&  L pol percent & V pol percent & $|$V$|$ pol percent &\multicolumn{1}{c}{$\Delta$ RM} & \multicolumn{1}{c}{$\Delta$ DM} &\multicolumn{1}{c}{Scat. time} & Scat. index \\
&  \multicolumn{1}{c}{(phase)}  & & & & & \multicolumn{1}{c}{(rad m$^{-2}$)} & \multicolumn{1}{c}{(pc cm$^{-2}$)} & \multicolumn{1}{c}{($\mu$s)} & \\
\hline
2 (ing) & 0.152 & 0.89 ( 5 ) & 42 ( 3 ) & 1 ( 3 ) & 13 ( 3 ) & 1.3 ( 8 ) & 0.0014 ( 7 ) &   &   \\
\hline
& 0.155 & 0.93 ( 5 ) & 39 ( 2 ) & 4 ( 3 ) & 20 ( 3 ) & 0.9 ( 16 ) & 0.0010 ( 5 ) &   &   \\
\hline
& 0.158 & 0.86 ( 4 ) & 52 ( 4 ) & 11 ( 4 ) & 17 ( 4 ) &  $-$0.1 ( 8 ) & 0.000 ( 1 ) &   &   \\
\hline
& 0.160 & 0.99 ( 4 ) & 44 ( 3 ) &  $-$5 ( 3 ) & 18 ( 3 ) & 1.3 ( 8 ) & 0.0013 ( 6 ) &   &   \\
\hline
& 0.163 & 1.04 ( 4 ) & 42 ( 3 ) & 5 ( 3 ) & 12 ( 3 ) &  $-$0.2 ( 7 ) & 0.0010 ( 4 ) &   &   \\
\hline
& 0.166 & 0.87 ( 4 ) & 43 ( 4 ) &  $-$6 ( 3 ) & 13 ( 3 ) & 1.2 ( 8 ) & 0.000 ( 1 ) &   &   \\
\hline
& 0.168 & 0.95 ( 4 ) & 44 ( 4 ) &  $-$1 ( 3 ) & 15 ( 3 ) & 0.1 ( 9 ) & 0.0016 ( 7 ) &   &   \\
\hline
& 0.171 & 0.93 ( 4 ) & 43 ( 4 ) &  $-$11 ( 4 ) & 15 ( 3 ) & 1.1 ( 8 ) & 0.001 ( 1 ) &   &   \\
\hline
& 0.174 & 0.97 ( 4 ) & 47 ( 4 ) & 9 ( 3 ) & 13 ( 3 ) & 0.5 ( 7 ) &  $-$0.0002 ( 8 ) &   &   \\
\hline
& 0.177 & 0.89 ( 5 ) & 34 ( 4 ) &  $-$4 ( 3 ) & 14 ( 2 ) & 4.3 ( 10 ) & 0.0030 ( 8 ) &   &   \\
\hline
& 0.179 & 0.83 ( 5 ) & 27 ( 4 ) &  $-$2 ( 3 ) & 24 ( 3 ) &   & 0.0005 ( 10 ) &   &   \\
\hline
& 0.182 & 0.84 ( 5 ) & 20 ( 5 ) & 2 ( 3 ) & 15 ( 3 ) &   & 0.0033 ( 7 ) &   &   \\
\hline
& 0.185 & 0.82 ( 5 ) & 16 ( 3 ) &  $-$6 ( 3 ) & 25 ( 3 ) &   & 0.0050 ( 9 ) &   &   \\
\hline
& 0.188 & 0.85 ( 4 ) & 35 ( 5 ) &  $-$3 ( 3 ) & 21 ( 3 ) &   & 0.002 ( 1 ) &   &   \\
\hline
& 0.190 & 0.91 ( 4 ) & 19 ( 2 ) &  $-$7 ( 4 ) & 15 ( 3 ) &   & 0.015 ( 1 ) & 29 ( 4 ) &  $-$4.5 ( 10 ) \\
\hline
& 0.193 & 0.71 ( 4 ) & 34 ( 5 ) & 4 ( 4 ) & 16 ( 4 ) &   & 0.029 ( 4 ) & 177 ( 22 ) &  $-$1.2 ( 7 ) \\
\hline
\end{tabular}
\end{table*}

\begin{table*}
\centering 
\caption{Same as Table \ref{tab:data_26U1} for the egress of eclipse \#2 using 32s sub-integrations.}
\label{tab:data_26U2}
\centering
\renewcommand{\arraystretch}{1.0}
\vskip 0.1cm
\begin{tabular}{c|c|c|c|c|c|c|c|c|c}
\hline
Eclipse \#   & \multicolumn{1}{c}{orbital} & I (corr)&  L pol percent & V pol percent & $|$V$|$ pol percent &\multicolumn{1}{c}{$\Delta$ RM} & \multicolumn{1}{c}{$\Delta$ DM} &\multicolumn{1}{c}{Scat. time} & Scat. index \\
&  \multicolumn{1}{c}{(phase)}  & & & & & \multicolumn{1}{c}{(rad m$^{-2}$)} & \multicolumn{1}{c}{(pc cm$^{-2}$)} & \multicolumn{1}{c}{($\mu$s)} & \\
\hline
2 (eg) & 0.310 & 0.47 ( 5 ) & 26 ( 6 ) &  $-$1 ( 7 ) & 21 ( 6 ) &   & 0.101 ( 6 ) & 926 ( 136 ) &  $-$4.2 ( 18 ) \\
\hline
& 0.313 & 0.87 ( 5 ) & 19 ( 1 ) &  $-$4 ( 4 ) & 14 ( 3 ) &   & 0.066 ( 4 ) & 309 ( 26 ) &  $-$3.2 ( 9 ) \\
\hline
& 0.316 & 0.91 ( 4 ) & 23 ( 3 ) & 1 ( 3 ) & 15 ( 3 ) &   & 0.049 ( 2 ) & 130 ( 11 ) &  $-$2.4 ( 5 ) \\
\hline
& 0.318 & 1.11 ( 5 ) & 17 ( 3 ) & 9 ( 3 ) & 12 ( 3 ) &   & 0.035 ( 1 ) & 35 ( 4 ) &  $-$5.3 ( 7 ) \\
\hline
& 0.321 & 1.11 ( 5 ) & 11 ( 3 ) & 5 ( 2 ) & 13 ( 2 ) &   & 0.034 ( 1 ) & 39 ( 4 ) &  $-$3.1 ( 6 ) \\
\hline
& 0.324 & 1.08 ( 4 ) & 16 ( 3 ) &  $-$4 ( 3 ) & 14 ( 2 ) &   & 0.024 ( 1 ) & 51 ( 4 ) &  $-$3.2 ( 5 ) \\
\hline
& 0.326 & 1.01 ( 5 ) & 17 ( 3 ) &  $-$8 ( 3 ) & 14 ( 3 ) &   & 0.0180 ( 8 ) & 24 ( 3 ) &  $-$3.3 ( 7 ) \\
\hline
& 0.329 & 1.12 ( 5 ) & 13 ( 3 ) & 3 ( 2 ) & 15 ( 2 ) &   & 0.0142 ( 6 ) &   &   \\
\hline
& 0.332 & 1.07 ( 5 ) & 13 ( 4 ) & 5 ( 3 ) & 13 ( 3 ) &   & 0.0141 ( 5 ) &   &   \\
\hline
& 0.335 & 1.02 ( 5 ) & 13 ( 3 ) & 0 ( 3 ) & 14 ( 3 ) &   & 0.0089 ( 9 ) &   &   \\
\hline
& 0.337 & 1.14 ( 4 ) & 17 ( 4 ) & 5 ( 3 ) & 11 ( 3 ) &   & 0.013 ( 1 ) &   &   \\
\hline
& 0.340 & 1.14 ( 4 ) & 14 ( 4 ) & 5 ( 2 ) & 14 ( 2 ) &   & 0.0146 ( 6 ) &   &   \\
\hline
& 0.343 & 1.02 ( 4 ) & 18 ( 4 ) & 3 ( 3 ) & 23 ( 3 ) &   & 0.0092 ( 8 ) &   &   \\
\hline
& 0.346 & 1.07 ( 4 ) & 22 ( 3 ) & 9 ( 2 ) & 20 ( 2 ) &   & 0.0067 ( 6 ) &   &   \\
\hline
& 0.348 & 1.30 ( 4 ) & 15 ( 3 ) & 1 ( 2 ) & 14 ( 2 ) &   & 0.0084 ( 5 ) &   &   \\
\hline
& 0.351 & 1.01 ( 5 ) & 9 ( 3 ) &  $-$3 ( 3 ) & 18 ( 3 ) &   & 0.0051 ( 9 ) &   &   \\
\hline
& 0.354 & 1.07 ( 4 ) & 17 ( 3 ) &  $-$0 ( 3 ) & 14 ( 2 ) & 2.6 ( 13 ) & 0.0012 ( 7 ) &   &   \\
\hline
& 0.356 & 1.22 ( 5 ) & 31 ( 2 ) & 7 ( 2 ) & 17 ( 2 ) & 0.5 ( 7 ) & 0.0006 ( 7 ) &   &   \\
\hline
& 0.359 & 1.36 ( 4 ) & 35 ( 3 ) &  $-$3 ( 2 ) & 14 ( 2 ) & 1.0 ( 6 ) & 0.0009 ( 9 ) &   &   \\
\hline
& 0.362 & 1.27 ( 5 ) & 38 ( 2 ) & 2 ( 2 ) & 11 ( 2 ) &  $-$0.6 ( 6 ) & 0.0007 ( 7 ) &   &   \\
\hline
& 0.365 & 1.14 ( 4 ) & 55 ( 3 ) &  $-$3 ( 2 ) & 13 ( 2 ) & 0.4 ( 5 ) & 0.0003 ( 9 ) &   &   \\
\hline
& 0.367 & 1.32 ( 5 ) & 41 ( 3 ) & 5 ( 2 ) & 12 ( 2 ) &  $-$0.5 ( 5 ) & 0.0012 ( 7 ) &   &   \\
\hline
& 0.370 & 1.15 ( 4 ) & 51 ( 3 ) & 2 ( 2 ) & 13 ( 2 ) & 0.4 ( 5 ) & 0.001 ( 1 ) &   &   \\
\hline
& 0.373 & 1.25 ( 5 ) & 35 ( 2 ) & 4 ( 2 ) & 13 ( 2 ) & 1.1 ( 7 ) & 0.0010 ( 8 ) &   &   \\
\hline
& 0.375 & 1.25 ( 4 ) & 46 ( 3 ) &  $-$3 ( 2 ) & 16 ( 2 ) &  $-$0.3 ( 6 ) & 0.000 ( 1 ) &   &   \\
\hline
& 0.378 & 1.13 ( 4 ) & 54 ( 3 ) &  $-$3 ( 2 ) & 11 ( 2 ) &  $-$0.3 ( 5 ) & 0.0016 ( 7 ) &   &   \\
\hline
& 0.381 & 1.20 ( 4 ) & 49 ( 3 ) & 1 ( 2 ) & 17 ( 2 ) & 1.2 ( 5 ) & 0.0029 ( 6 ) &   &   \\
\hline
& 0.384 & 1.17 ( 5 ) & 43 ( 3 ) &  $-$3 ( 2 ) & 12 ( 2 ) & 0.2 ( 5 ) & 0.0001 ( 7 ) &   &   \\
\hline
& 0.386 & 1.24 ( 4 ) & 44 ( 2 ) & 3 ( 2 ) & 10 ( 2 ) &  $-$0.7 ( 5 ) & 0.0019 ( 6 ) &   &   \\
\hline
& 0.389 & 1.33 ( 4 ) & 38 ( 2 ) &  $-$6 ( 2 ) & 13 ( 2 ) & 0.2 ( 5 ) & 0.0014 ( 6 ) &   &   \\
\hline
& 0.392 & 1.22 ( 5 ) & 40 ( 2 ) & 2 ( 2 ) & 18 ( 2 ) & 0.1 ( 6 ) & 0.0013 ( 7 ) &   &   \\
\hline
& 0.395 & 1.21 ( 4 ) & 50 ( 3 ) & 4 ( 2 ) & 15 ( 2 ) &  $-$0.1 ( 7 ) & 0.0026 ( 7 ) &   &   \\
\hline
& 0.397 & 1.25 ( 4 ) & 46 ( 3 ) & 5 ( 3 ) & 11 ( 3 ) & 0.9 ( 6 ) & 0.0013 ( 3 ) &   &   \\
\hline

\end{tabular}
\end{table*}

\begin{table*}
\centering 
\caption{Same as Table \ref{tab:data_26U1} for eclipse \#3.}
\label{tab:data_26U3}
\centering
\renewcommand{\arraystretch}{1.0}
\vskip 0.1cm
\begin{tabular}{c|c|c|c|c|c|c|c|c|c}
\hline
Eclipse \#   & \multicolumn{1}{c}{orbital} & I (corr)&  L pol percent & V pol percent & $|$V$|$ pol percent &\multicolumn{1}{c}{$\Delta$ RM} & \multicolumn{1}{c}{$\Delta$ DM} &\multicolumn{1}{c}{Scat. time} & Scat. index \\
&  \multicolumn{1}{c}{(phase)}  & & & & & \multicolumn{1}{c}{(rad m$^{-2}$)} & \multicolumn{1}{c}{(pc cm$^{-2}$)} & \multicolumn{1}{c}{($\mu$s)} & \\
\hline
3 (ing) & 0.150 & 0.97 ( 6 ) & 39 ( 5 ) & $-$6 ( 4 ) & 12 ( 4 ) & 0 ( 1 ) & 0.0011 ( 6 ) &   &   \\
\hline
& 0.156 & 0.76 ( 6 ) & 43 ( 5 ) & 6 ( 4 ) & 19 ( 4 ) &   & 0.0006 ( 4 ) &   &   \\
\hline
& 0.161 & 0.79 ( 5 ) & 61 ( 5 ) & 14 ( 4 ) & 14 ( 4 ) & $-$1 ( 1 ) & 0.001 ( 1 ) &   &   \\
\hline
& 0.167 & 0.91 ( 6 ) & 40 ( 5 ) & 3 ( 4 ) & 14 ( 4 ) &   & 0.0012 ( 4 ) &   &   \\
\hline
& 0.172 & 0.73 ( 6 ) & 59 ( 6 ) & 6 ( 5 ) & 13 ( 4 ) & $-$4 ( 2 ) & 0.0019 ( 7 ) &   &   \\
\hline
& 0.178 & 0.74 ( 5 ) & 49 ( 7 ) & $-$10 ( 4 ) & 25 ( 4 ) &   & $-$0.0008 ( 4 ) &   &   \\
\hline
& 0.183 & 0.82 ( 5 ) & 37 ( 5 ) & $-$5 ( 4 ) & 19 ( 4 ) &   & 0.001 ( 1 ) &   &   \\
\hline
& 0.189 & 0.70 ( 6 ) & 8 ( 7 ) & 3 ( 3 ) & 17 ( 3 ) &   & $-$0.001 ( 2 ) & 76 ( 10 ) & $-$4.1 ( 7 ) \\
\hline
\hline
3 (eg) & 0.319 & 0.65 ( 5 ) & 50 ( 7 ) & 0 ( 5 ) & 17 ( 4 ) &   & 0.035 ( 3 ) & 179 ( 25 ) & $-$4.3 ( 7 ) \\
\hline
& 0.325 & 0.67 ( 5 ) & 37 ( 5 ) & 11 ( 5 ) & 25 ( 4 ) &   & 0.024 ( 2 ) & 49 ( 8 ) & $-$3 ( 1 ) \\
\hline
& 0.330 & 0.68 ( 6 ) & 25 ( 6 ) & 8 ( 4 ) & 35 ( 4 ) &   & 0.012 ( 2 ) &   &   \\
\hline
& 0.336 & 0.59 ( 5 ) & 39 ( 8 ) & 5 ( 7 ) & 22 ( 6 ) &   & 0.002 ( 2 ) & 33 ( 5 ) & $-$2 ( 1 ) \\
\hline
& 0.341 & 0.51 ( 5 ) & 55 ( 10 ) & $-$2 ( 7 ) & 22 ( 6 ) &   & 0.003 ( 1 ) &   &   \\
\hline
& 0.347 & 0.71 ( 5 ) & 32 ( 7 ) & 2 ( 4 ) & 29 ( 4 ) &   & 0.002 ( 1 ) &   &   \\
\hline
& 0.352 & 0.78 ( 6 ) & 13 ( 5 ) & $-$4 ( 4 ) & 18 ( 4 ) &   & 0.0004 ( 2 ) &   &   \\
\hline
& 0.357 & 0.59 ( 5 ) & 73 ( 8 ) & $-$3 ( 6 ) & 17 ( 6 ) &   & 0.0014 ( 4 ) &   &   \\
\hline
& 0.363 & 0.95 ( 6 ) & 31 ( 4 ) & $-$2 ( 3 ) & 14 ( 3 ) &   & 0.0006 ( 7 ) &   &   \\
\hline
& 0.368 & 0.68 ( 6 ) & 76 ( 7 ) & $-$1 ( 6 ) & 11 ( 5 ) & 2 ( 2 ) & 0.0009 ( 7 ) &   &   \\
\hline
& 0.374 & 0.92 ( 5 ) & 44 ( 7 ) & 11 ( 3 ) & 25 ( 3 ) &   & 0.0019 ( 4 ) &   &   \\
\hline
& 0.379 & 0.80 ( 6 ) & 34 ( 5 ) & 6 ( 4 ) & 18 ( 3 ) &   & 0.0011 ( 4 ) &   &   \\
\hline
& 0.385 & 0.77 ( 5 ) & 69 ( 5 ) & 15 ( 5 ) & 20 ( 4 ) & 2 ( 1 ) & 0.0014 ( 6 ) &   &   \\
\hline
& 0.390 & 0.70 ( 5 ) & 64 ( 5 ) & 16 ( 5 ) & 27 ( 4 ) &   & 0.0017 ( 5 ) &   &   \\
\hline
& 0.396 & 0.78 ( 6 ) & 48 ( 4 ) & 2 ( 4 ) & 23 ( 4 ) & $-$3 ( 2 ) & $-$0.001 ( 1 ) &   &   \\
\hline
\end{tabular}
\end{table*}

\begin{table*}
\centering 
\caption{Same as Table \ref{tab:data_26U1} for eclipse \#4.}
\label{tab:data_26U4}
\centering
\renewcommand{\arraystretch}{1.0}
\vskip 0.1cm
\begin{tabular}{c|c|c|c|c|c|c|c|c|c}
\hline
Eclipse \#   & \multicolumn{1}{c}{orbital} & I (corr)&  L pol percent & V pol percent & $|$V$|$ pol percent &\multicolumn{1}{c}{$\Delta$ RM} & \multicolumn{1}{c}{$\Delta$ DM} &\multicolumn{1}{c}{Scat. time} & Scat. index \\
&  \multicolumn{1}{c}{(phase)}  & & & & & \multicolumn{1}{c}{(rad m$^{-2}$)} & \multicolumn{1}{c}{(pc cm$^{-2}$)} & \multicolumn{1}{c}{($\mu$s)} & \\
\hline
4 (ing) & 0.151 & 1.17 ( 6 ) & 56 ( 6 ) & 8 ( 4 ) & 16 ( 3 ) & 1.3 ( 19 ) & 0.0006 ( 5 ) &   &   \\
\hline
& 0.157 & 1.12 ( 8 ) & 35 ( 5 ) & 0 ( 4 ) & 16 ( 4 ) &   & 0.0010 ( 9 ) &   &   \\
\hline
& 0.162 & 1.07 ( 7 ) & 61 ( 6 ) & 0 ( 4 ) & 8 ( 4 ) & 4 ( 2 ) & $-$0.0001 ( 8 ) &   &   \\
\hline
& 0.168 & 1.27 ( 7 ) & 52 ( 5 ) & $-$2 ( 4 ) & 21 ( 3 ) & 2 ( 1 ) & 0.0007 ( 12 ) &   &   \\
\hline
& 0.173 & 1.33 ( 8 ) & 32 ( 4 ) & 3 ( 3 ) & 20 ( 3 ) & 0 ( 2 ) & 0.0015 ( 7 ) &   &   \\
\hline
& 0.179 & 1.20 ( 8 ) & 12 ( 5 ) & $-$2 ( 4 ) & 15 ( 4 ) &   & 0.002 ( 2 ) &   &   \\
\hline
& 0.184 & 0.99 ( 7 ) & 27 ( 6 ) & $-$7 ( 4 ) & 20 ( 4 ) &   & 0.002 ( 1 ) &   &   \\
\hline
& 0.190 & 0.84 ( 7 ) & 32 ( 6 ) & 6 ( 5 ) & 24 ( 5 ) &   & 0.011 ( 3 ) & 165 ( 16 ) & $-$0.7 ( 8 ) \\
\hline
\hline
4 (eg) & 0.315 & 0.83 ( 7 ) & 34 ( 7 ) & 0 ( 4 ) & 27 ( 4 ) &   & 0.091 ( 4 ) & 182 ( 24 ) & 0.0 ( 10 ) \\
\hline
& 0.320 & 0.73 ( 8 ) & 29 ( 8 ) & $-$5 ( 5 ) & 36 ( 5 ) &   & 0.018 ( 3 ) & 99 ( 13 ) & $-$3.3 ( 11 ) \\
\hline
& 0.326 & 0.93 ( 7 ) & 51 ( 7 ) & $-$9 ( 5 ) & 24 ( 5 ) &   & 0.022 ( 2 ) &   &   \\
\hline
& 0.331 & 1.14 ( 7 ) & 26 ( 4 ) & 4 ( 4 ) & 18 ( 4 ) &   & 0.018 ( 2 ) &   &   \\
\hline
& 0.337 & 1.00 ( 6 ) & 47 ( 6 ) & 3 ( 5 ) & 25 ( 4 ) &   & 0.016 ( 2 ) &   &   \\
\hline
& 0.342 & 0.95 ( 7 ) & 20 ( 5 ) & $-$4 ( 5 ) & 18 ( 4 ) &   & 0.011 ( 2 ) & 44 ( 6 ) & $-$4.8 ( 11 ) \\
\hline
& 0.347 & 1.06 ( 7 ) & 28 ( 6 ) & 6 ( 4 ) & 18 ( 3 ) &   & 0.007 ( 2 ) & 60 ( 7 ) & $-$4.4 ( 8 ) \\
\hline
& 0.353 & 0.99 ( 7 ) & 17 ( 6 ) & 10 ( 4 ) & 21 ( 4 ) &   & $-$0.000 ( 1 ) &   &   \\
\hline
& 0.358 & 0.72 ( 7 ) & 30 ( 9 ) & 12 ( 8 ) & 19 ( 7 ) &   & 0.003 ( 1 ) &   &   \\
\hline
& 0.364 & 0.79 ( 6 ) & 24 ( 6 ) & 2 ( 4 ) & 32 ( 4 ) &   & 0.0023 ( 3 ) &   &   \\
\hline
& 0.369 & 0.94 ( 8 ) & 17 ( 3 ) & 11 ( 4 ) & 29 ( 4 ) &   & $-$0.002 ( 2 ) &   &   \\
\hline
& 0.375 & 0.97 ( 6 ) & 55 ( 6 ) & 5 ( 4 ) & 22 ( 4 ) & 3 ( 1 ) & 0.001 ( 1 ) &   &   \\
\hline
& 0.380 & 0.97 ( 7 ) & 44 ( 7 ) & $-$10 ( 4 ) & 26 ( 4 ) &   & 0.0016 ( 7 ) &   &   \\
\hline
& 0.386 & 0.92 ( 8 ) & 37 ( 6 ) & $-$13 ( 4 ) & 19 ( 4 ) & 6 ( 2 ) & 0.0004 ( 9 ) &   &   \\
\hline
& 0.391 & 0.96 ( 7 ) & 51 ( 7 ) & 3 ( 4 ) & 13 ( 4 ) & $-$1 ( 2 ) & 0.0007 ( 7 ) &   &   \\
\hline
& 0.397 & 1.21 ( 7 ) & 35 ( 5 ) & 8 ( 3 ) & 14 ( 3 ) & 3 ( 1 ) & 0.0005 ( 9 ) &   &   \\
\hline
\end{tabular}
\end{table*}

\begin{table*}
\centering 
\caption{Same as Table \ref{tab:data_26U1} for eclipse \#5.}
\label{tab:data_26U5}
\centering
\renewcommand{\arraystretch}{1.0}
\vskip 0.1cm
\begin{tabular}{c|c|c|c|c|c|c|c|c|c}
\hline
Eclipse \#   & \multicolumn{1}{c}{orbital} & I (corr)&  L pol percent & V pol percent & $|$V$|$ pol percent &\multicolumn{1}{c}{$\Delta$ RM} & \multicolumn{1}{c}{$\Delta$ DM} &\multicolumn{1}{c}{Scat. time} & Scat. index \\
&  \multicolumn{1}{c}{(phase)}  & & & & & \multicolumn{1}{c}{(rad m$^{-2}$)} & \multicolumn{1}{c}{(pc cm$^{-2}$)} & \multicolumn{1}{c}{($\mu$s)} & \\
\hline
5 (ing) & 0.152 & 1.04 ( 10 ) & 62 ( 7 ) & 7 ( 7 ) & 34 ( 6 ) &   & 0.0011 ( 7 ) &   &   \\
\hline
& 0.158 & 0.93 ( 11 ) & 59 ( 6 ) & $-$5 ( 8 ) & 22 ( 8 ) &   &  $-$0.001 ( 1 ) &   &   \\
\hline
& 0.163 & 1.14 ( 11 ) & 45 ( 6 ) &  $-$5 ( 6 ) & 19 ( 6 ) &   & 0.0017 ( 6 ) &   &   \\
\hline
& 0.169 & 0.95 ( 11 ) & 59 ( 9 ) &  $-$8 ( 8 ) & 20 ( 7 ) &   & 0.001 ( 2 ) &   &   \\
\hline
& 0.174 & 1.07 ( 11 ) & 53 ( 8 ) & 7 ( 5 ) & 18 ( 5 ) &   & 0.003 ( 2 ) &   &   \\
\hline
& 0.180 & 0.90 ( 9 ) & 65 ( 12 ) &  $-$8 ( 7 ) & 31 ( 7 ) &   &  $-$0.003 ( 1 ) &   &   \\
\hline
& 0.185 & 0.57 ( 12 ) & 0 ( 0 ) & 0 ( 0 ) & 0 ( 0 ) &   & 0.004 ( 2 ) &   &   \\
\hline
& 0.191 & 0.48 ( 9 ) & 101 ( 21 ) & 13 ( 17 ) & 31 ( 16 ) &   & 0.005 ( 3 ) & 34 ( 13 ) &  $-$3.3 ( 16 ) \\
\hline
\hline
5 (eg) & 0.332 & 0.91 ( 11 ) & 35 ( 8 ) & 1 ( 8 ) & 20 ( 8 ) &   & 0.008 ( 2 ) &   &   \\
\hline
& 0.338 & 1.31 ( 11 ) & 30 ( 6 ) & 8 ( 5 ) & 12 ( 5 ) &   & 0.006 ( 1 ) &   &   \\
\hline
& 0.343 & 1.15 ( 10 ) & 60 ( 8 ) &  $-$14 ( 6 ) & 29 ( 5 ) &   & 0.004 ( 1 ) &   &   \\
\hline
& 0.349 & 1.34 ( 10 ) & 33 ( 5 ) &  $-$8 ( 5 ) & 26 ( 4 ) &   & 0.0041 ( 8 ) &   &   \\
\hline
& 0.354 & 1.07 ( 11 ) & 40 ( 6 ) &  $-$2 ( 6 ) & 15 ( 6 ) &   & 0.001 ( 1 ) &   &   \\
\hline
& 0.359 & 1.18 ( 9 ) & 41 ( 7 ) & 9 ( 6 ) & 17 ( 5 ) &   & 0.0013 ( 7 ) &   &   \\
\hline
& 0.365 & 0.76 ( 11 ) & 24 ( 9 ) & 0 ( 10 ) & 16 ( 10 ) &   & 0.002 ( 2 ) &   &   \\
\hline
& 0.370 & 0.99 ( 11 ) & 42 ( 8 ) &  $-$13 ( 8 ) & 21 ( 8 ) &   & 0.0014 ( 6 ) &   &   \\
\hline
& 0.376 & 0.44 ( 11 ) & 0 ( 0 ) & 0 ( 0 ) & 0 ( 0 ) &   &  $-$0.003 ( 2 ) &   &   \\
\hline
& 0.381 & 0.62 ( 11 ) & 97 ( 17 ) & 3 ( 11 ) & 61 ( 10 ) &   & 0.0023 ( 9 ) &   &   \\
\hline
& 0.387 & 1.03 ( 10 ) & 46 ( 9 ) &  $-$6 ( 7 ) & 21 ( 7 ) &   & 0.0008 ( 5 ) &   &   \\
\hline
& 0.392 & 0.80 ( 10 ) & 67 ( 12 ) &  $-$8 ( 9 ) & 33 ( 9 ) &   & 0.0001 ( 8 ) &   &   \\
\hline
& 0.398 & 0.81 ( 12 ) & 54 ( 8 ) & 6 ( 7 ) & 18 ( 7 ) &   & 0.0008 ( 5 ) &   &   \\
\hline
\end{tabular}
\end{table*}

\begin{table*}
\centering 
\caption{Same as Table \ref{tab:data_26U1} for eclipse \#6.}
\label{tab:data_26U6}
\centering
\renewcommand{\arraystretch}{1.0}
\vskip 0.1cm
\begin{tabular}{c|c|c|c|c|c|c|c|c|c}
\hline
Eclipse \#   & \multicolumn{1}{c}{orbital} & I (corr)&  L pol percent & V pol percent & $|$V$|$ pol percent &\multicolumn{1}{c}{$\Delta$ RM} & \multicolumn{1}{c}{$\Delta$ DM} &\multicolumn{1}{c}{Scat. time} & Scat. index \\
&  \multicolumn{1}{c}{(phase)}  & & & & & \multicolumn{1}{c}{(rad m$^{-2}$)} & \multicolumn{1}{c}{(pc cm$^{-2}$)} & \multicolumn{1}{c}{($\mu$s)} & \\
\hline
6 (ing) & 0.152 & 0.93 ( 13 ) & 45 ( 12 ) & 4 ( 9 ) & 30 ( 8 ) &   & 0.0006 ( 5 ) &   &   \\
\hline
& 0.158 & 0.89 ( 15 ) & 32 ( 9 ) & 12 ( 9 ) & 41 ( 9 ) &   & 0.001 ( 1 ) &   &   \\
\hline
& 0.163 & 0.86 ( 12 ) & 88 ( 13 ) & $-$11 ( 8 ) & 39 ( 8 ) &   & 0.0020 ( 9 ) &   &   \\
\hline
& 0.168 & 0.79 ( 12 ) & 79 ( 10 ) & $-$1 ( 10 ) & 50 ( 9 ) &   & 0.0011 ( 7 ) &   &   \\
\hline
& 0.174 & 0.69 ( 13 ) & 50 ( 15 ) & $-$6 ( 9 ) & 25 ( 9 ) &   & 0.002 ( 2 ) &   &   \\
\hline
& 0.179 & 0.98 ( 12 ) & 58 ( 13 ) & 6 ( 9 ) & 23 ( 8 ) &   & 0.0017 ( 8 ) &   &   \\
\hline
\end{tabular}
\end{table*}


\bsp	
\label{lastpage}
\end{document}